# Emergence of Animals from Heat Engines. Part 1. Before the Snowball Earths

A.W.J. Muller
*Swammerdam Institute for Life Sciences*
*University of Amsterdam, Kruislaan 318, 1098 SM Amsterdam, The Netherlands*
a.w.j.muller @ uva.nl

Tel.+ 31 20 525 5187 Fax : 31 20 525 7934

**Abstract**
Previous studies modelled the origin of life and the emergence of photosynthesis on the early Earth— i.e. the origin of *plants* —in terms of biological heat engines that worked on thermal cycling caused by suspension in convecting water. In this new series of studies, heat engines using a more complex mechanism for thermal cycling are invoked to explain the origin of *animals* as well. Biological exploitation of the thermal gradient above a submarine hydrothermal vent is hypothesized, where a relaxation oscillation in the length of a protein 'thermotether' would have yielded the thermal cycling required for thermosynthesis. Such a thermal transition driven movement is not impeded by the low Reynolds number of a small scale. In the model the thermotether together with the protein export apparatus evolved into a 'flagellar proton pump' that turned into today's bacterial flagellar motor after the acquisition of the proton-pumping respiratory chain. The flagellar pump resembles Feynman's ratchet, and the 'flagellar computer' that implements chemotaxis a Turing machine: the stator would have functioned as Turing's paper tape and the stator's proton-transferring subunits with their variable conformation as the symbols on the tape. The existence of a cellular control centre in the cilium of the eukaryotic cell is proposed that would have evolved from the prokaryotic flagellar computer.

Keywords: flagellar motor, thermosynthesis, Reynolds Number, Feynman's ratchet, Turing machine

Abbreviations:
FP: First Protein, a protein that can perform condensation reactions, especially ATP synthesis, during thermal cycling
TSC: thermosynthesis working on thermal cycling caused by convection,
TSG: thermosynthesis working on a thermal gradient

## 1. Introduction

Thermosynthesis involves the idea of biological heat engines that work on thermal cycling (Fig. 1). In previous studies the notion was applied in models for the origin of life and the origin of photosynthesis (Muller, 1983, 1985, 1993, 1995, 2005a, 2005b; Muller and Schulze-Makuch, 2006a, 2006b), i.e. the origin of plants. In these studies the required thermal cycling was the effect of suspension in convecting water such as present in volcanic hot springs.

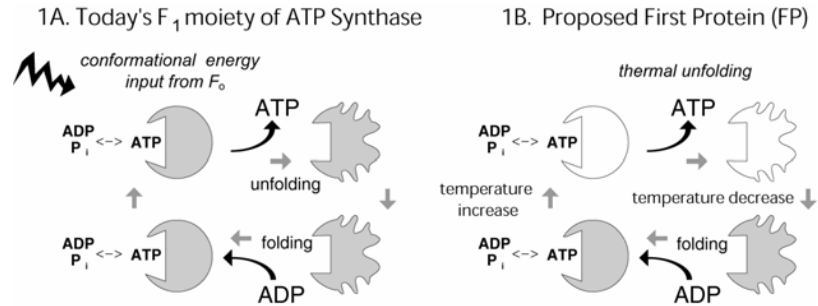

Fig. 1. *The previously proposed thermosynthesis mechanism based on a thermal variation of the binding change mechanism* During chemiosmosis—a partial process of both respiration and photosynthesis—the key biological energy carrier ATP is generated inside the $F_1$ moiety of the ATP Synthase enzyme (1A). Bound ADP and phosphate form tightly bound ATP. When protons cross the membrane from high to low electric potential in the $F_o$ moiety of ATP Synthase, their free energy is transduced (crooked arrow) to the $F_1$ moiety of ATP Synthase, causing the bound ATP to be released. It is proposed that in the first protein (FP) the release of similarly formed bound ATP (1B) was effected by thermal unfolding. The thermosynthesis mechanism works on thermal cycling.

Large multicellular animals (metazoans) suddenly emerge in the fossil record of the late Proterozoic (Conway Morris, 2006 ; Darwin, 1859; Glaessner, 1984; McCall, 2006; Narbonne, 1998; Xiao, 2005), right after a series of extensive glaciations, 'Snowball Earths' (Hoffman et al. 1998; Kirschvink, 1992; Mawson, 1949) that date from 750 - 550 Ma (Million years ago). Here we present the first in a series of studies that explains the origin of animals by thermosynthesis as well.

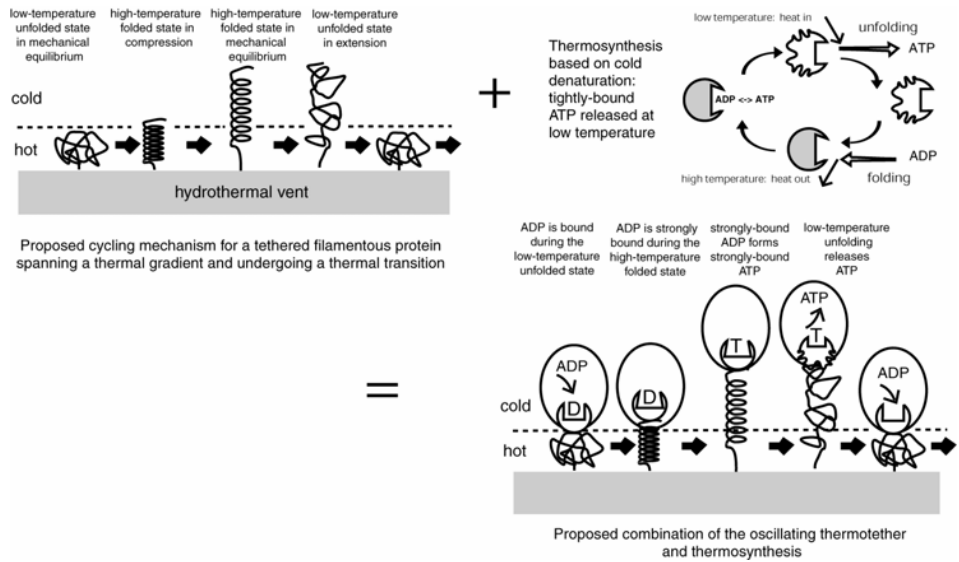

Fig. 2. *A thermotether that works on a relaxation oscillation in a thermal gradient is combined with the First Protein, yielding an ATP generator that works on a thermal gradient* In the thermal gradient above a hydrothermal vent, a thermotether performs a relaxation oscillation: while warm it folds, but as the relaxed folded and relaxed unfolded protein have different shapes, the new conformation is under strain. It cannot instantaneously relax to the new equilibrium shape. After relaxation by expansion, the thermotether comes in contact with the cold ocean water and cools. Cold denaturation sets in, which changes the conformation. The protein is again first in strain, and then relaxes, this time by contraction. An attached FP that works on cold denaturation (instead of on the hot denaturation process shown in Fig. 1) then permits the synthesis of ATP from the thermal cycling caused by the oscillating thermotether.

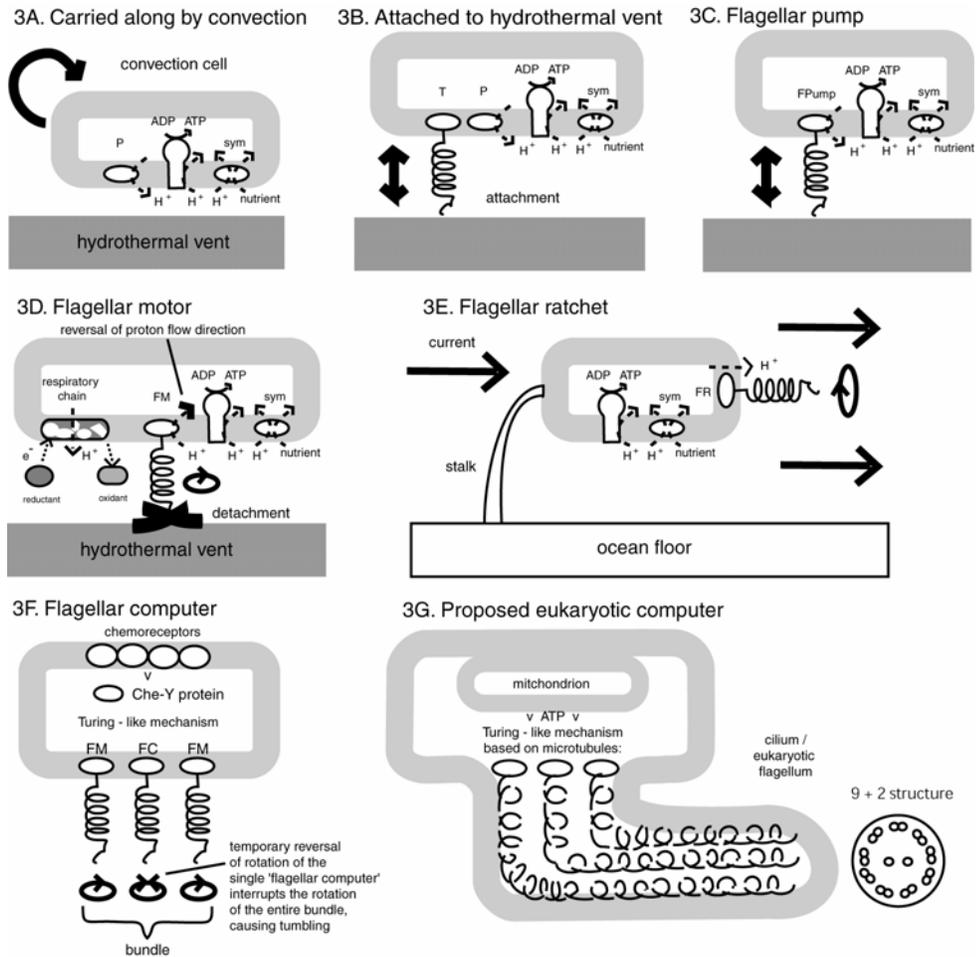

Fig. 3. *Overview of the proposed evolution of flagellum-based heat engines that worked on a thermal gradient*

3A. In a bacterium carried along by convection above a hydrothermal vent, a proton pump P was driven by the thermal cycling (Muller, 1995). The resulting proton gradient drove ATP Synthase and the uptake of nutrients by $H^+$/nutrient symport (sym). Since the mechanism does not require sunlight, it could have worked during a global glaciation.

3B. A filamentous protein T was attached to a hydrothermal vent, which permitted thermal cycling caused by temperature fluctuations on the surface of a hydrothermal vent. By turning into a thermotether, the filamentous protein permitted a free energy gain from a constant vent temperature as well.

3C. The proton pump combined with the thermotether to form a flagellar pump (FPump).

3D. After the end of a global glaciation, addition of the respiratory chain to the bacterial membrane enabled proton pumping by respiration while electrons moved from reductant to oxidant. Following disconnection of the flagellum from the hydrothermal vent, the flagellar pump functioned as flagellar motor (FM) while the proton flow through the flagellar pump reversed.

3E. Proton pumping by the flagellar ratchet (FR) was enabled by (1) a thermal difference between the flagellum and the stator in the flagellar motor, but also mechanical movement of the flagellum due to (2) non-thermal fluctuations or even (3) a steady flow.

3F. Emergence of today's bacterial flagellar computer (FC) in a bundle of flagella. One flagellum temporarily interrupts the cooperative functioning of a bundle, causing 'tumbling', a random change in direction. The frequency of tumbling is determined by chemoreception; this flagellar motor resembles the Turing machine.

3G. Instead of using protons, the eukaryotic flagellum works on ATP. It emerged upon symbiosis of (1) an ATP-yielding bacterium with (2) a host that descended from a flagellated bacterium in which the flagellar bundle had acquired an enclosing membrane and the flagella had been replaced by microtubules. The data processing capability of the eukaryotic cell (the 'eukaryotic computer') descended from the bacterial flagellar computer.

The thermosynthesizing ancestors of the animals would have lived, not on convection, but on the thermal gradient between cold ocean water and hot submarine hydrothermal vents (Muller, 2008). These ancestors may have done so rather early, in the Archean, 3235 Ma, since fossils from that time already suggest the occurrence of protein conformational changes driven by a thermal gradient (Rasmussen, 2000). It is hypothesized that shape oscillations sustained ATP synthesis at a time of a global glaciation. The first cell organelle with this capability was a protein 'thermotether' that in the thermal gradient performed a relaxation oscillation (Van der Pol and Van der Mark, 1928) caused by reversible low-temperature (~ 0 °C) unfolding, also called 'cold denaturation' (Fig. 2) (Brandts, 1967; Todgham et al. 2007). Merger of parts of ATP Synthase with the type-III protein export complex (Hueck, 1998) then yielded a flagellar proton pump (Fig. 3) that resembled a Feynman ratchet (Feynman et al. 1963; Machura, 2006; Von Smoluchowski, 1912). The bacterial flagellar motor (Berg, 2003) evolved from the flagellar pump when it reversed at the end of a global glaciation, after protons pumped by respiration had again become available. We end by considering the emergence of computers similar to Turing machines (Turing, 1937) from flagellar motors.

Thermosynthesis draws heavily on chemiosmosis, the ubiquitous mechanism of ATP synthesis (Cramer and Knaff, 1991; Dimroth et al. 2006; Mitchell, 1961, 1979; Williams, 1961). Although widely used in explaining physiological phenomena, application of chemiosmosis to the modelling of evolution and the origin of life is still rare (Harold, 2001).

Here we give a general introduction to a series of papers on the emergence of animals. We moreover consider thermosynthesis at hydrothermal vents early in the Proterozoic. The following studies will consider (1) the emergence of the Ediacaran organisms at hydrothermal vents during the late Proterozoic, (2) the Cambrian explosion, and (3) physiological effects of the temperature in today's animals.

## 2. Background

### 2.1 Simple Heat Engines

In physics and engineering the notion of a heat engine is well established. Many types have been described. The working substance can be a solid, liquid, vapour or a gas (Carnot, 1824). In addition to the familiar steam engine and the somewhat less familiar Stirling gas engine, heat engines have been constructed or proposed based on many principles, including temperature-dependent electric (Gonzalo, 1976; Hoh, 1963; Pulvari and Garcia, 1978; Sklar, 2005) and magnetic polarization (Tesla, 1889), and shape memory metal (Wekjira, 2001). Rubber is derived from a biological material produced by plants. It contracts with the temperature (Wiegand and Snyder, 1934) and heat engines based on rubber (Mullen et al., 1975; Savarino and Fisch, 1991; Singh, 1989) are of interest since proteins such as collagen, fibroin and elastin contract in the same way (Flory, 1953, 1956; Wöhlisch, 1940). More generally, almost any temperature-dependent material characteristic seems applicable in a heat engine.

Heat engines such as the gas turbines that sustain jet propulsion and drive electric power generators are complex in their construction of advanced materials (Reed, 2006) because of the benefit of closely approaching, at maximal temperature difference, the optimal Carnot efficiency. Complexity is, however, not inherent to heat engines. The convection cell, in which the rotating fluid does mechanical work, demonstrates that a heat engine can even be self-organizing. Convection is a ubiquitous and important process. Ultimately, most of the order in the Universe is derived from convection. Carnot (1824) already observed:

To heat . . . are due the vast movements which take place on the Earth,

and convection indeed sustains the weather in the atmosphere, the ocean currents in the hydrosphere, the plate tectonics in the lithosphere, and may sustain the geodynamo in the Earth's core that generates the Earth's magnetic field (Hollerbach, 1996; Roberts and Soward, 1992; Zhang and Schubert, 2000).

The previously proposed thermosynthesis processes invoked to explain the origin of life made use of convection in volcanic hot springs. The thermosynthesis processes considered in this study and the following ones are based on relaxation oscillations in the thermal gradient between a submarine hydrothermal vent and the ocean. This gradient derives from two grand convection cells. The heat source, the vent, is the result of the large convection cells in the Earth's mantle that drive plate tectonics (Schubert et al., 2001), and the heat sink, the cold ocean water, results from the thermohaline circulation in the ocean (Warren, 1981). In a large convection cell, warm ocean currents move along an erratic wind-driven path (Wunsch and Ferrari, 2004) along the ocean's surface from the tropics to convergence zones near the poles where the water cools, becomes more dense, sinks, and returns along the ocean floor—passing by the hydrothermal vents—to the tropics where upwelling occurs.

Simple but inefficient heat engines are of little interest to engineering but may still be of use in biology. A low efficiency is not a drawback when the heat source is freely available low grade heat. Since the middle of the nineteenth century there have been several proposals for the application of heat engine mechanisms to the life sciences (Broda, 1975; Florkin, 1972; Komissarov, 1990; Van Holde, 1980), especially for muscle (Engelmann, 1893). For this organ the application fails (Fick, 1893a, 1893b). The absence of detectable thermal gradients in muscle is incompatible with a heat engine mechanism given the muscle's known high efficiency. This incompatibility does not, however, contradict the *emergence* of muscle or of other organs from a heat engine.

In physics a heat engine is defined as a system that performs mechanical work with simultaneous heat transport from high to low temperature. In this study a system is also called a heat engine when heat transport down a thermal gradient is accompanied by the generation of a high free energy substance such as ATP.

Work is readily interchanged with kinetic energy. A relaxation oscillation can be driven by a change in temperature that causes a shape change that disconnects the thermal contact that caused the temperature change in the first place (Fig. 4). Bimetal is a combination of (1) invar, an alloy (FeNi36) that does not expand with the temperature (Guillaume, 1897), and (2) an alloy that does. Car signalling lights are an application of a similar relaxation oscillation. Light emission stops when an electrical current is interrupted by an electrical contact broken by a bending of the bimetal, the result of the temperature increase caused by the current: the current switches itself off after it was switched on.

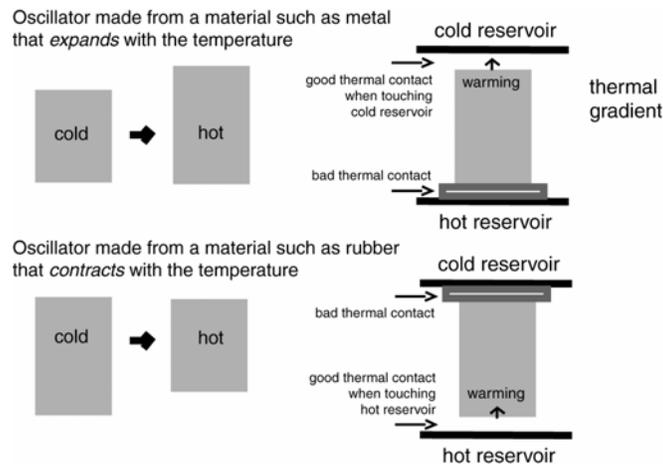

Fig. 4. *Relaxation oscillations in a thermal gradient based on shape changes with the temperature* An expansion due to a temperature change results in a thermal contact that in turn opposes this temperature change.

## 2.2 The Chemiosmotic Mechanism

Are biological heat engines possible? If a biological heat engine could be imagined, the well known adaptability of life, as shown by its versatility in using many *chemical* energy sources (Shock, 1990), would make it plausible that at least a few organisms could make use of this *physical* energy source as well.

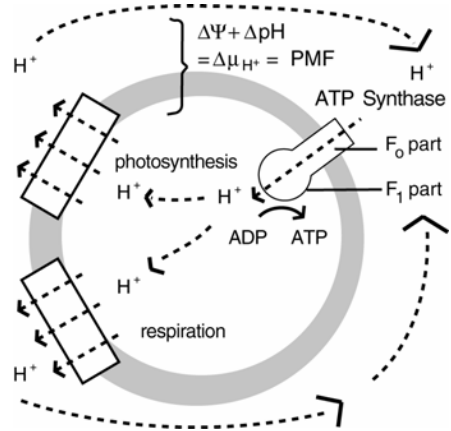

Fig. 5. *Basic principles of chemiosmosis*  During both photosynthesis and respiration protons are pumped across a membrane, which stores the free energy generated. The ATP Synthase enzyme transduces the free energy of this Proton Motive Force (PMF) into the free energy of ATP (Fig. 1).

In life, the ATP molecule is the energy currency. Almost all ATP is generated by the chemiosmotic mechanism, the biological energy conversion process that sustains both respiration and photosynthesis (Mitchell, 1961, 1979; Williams, 1961). Its machinery consists of enzymes and a biomembrane. The electrical potential difference across this biomembrane is part of the energy transduction process (Fig. 5). During respiration, electrons move from a reductant to an oxidant while simultaneously protons are pumped across a membrane; during photosynthesis, electrons excited by light also drive proton pumping. The free energy of protons at the high electrical membrane potential is transduced by the ATP Synthase enzyme into ATP while the pumped protons fall back through the membrane to the lower potential.

The $\Delta G_{0\ ATP}$ of ATP synthesis is 30 kJ/mol (Rosing, and Slater, 1972), and depends strongly on the pH and concentration of $Mg^{2+}$, ADP and ATP, which themselves are also highly variable. *In vivo* $\Delta G_{ATP}$ therefore can also vary strongly, and can be as high as 50 kJ/mol (Giersch et al., 1980). These values correspond to electrical membrane potential differences of

$$\Delta\Psi_1 = \Delta G_{ATP} / F = 310, \text{ resp. } 520 \text{ mV},$$

where $F$ is Faraday's constant. A value that we will use for numerical estimates is 360 mV (Schlodder et al., 1982). A difference in pH across the membrane, $\Delta pH$, can drive ATP synthesis too (a process driven by a difference in pH is endothermic (Hardt, 1980), just as the entropy rising step in the T-S cycle of a heat engine). A combination of $\Delta pH$ and $\Delta\Psi$ also works, and is called the 'electrochemical proton gradient across the membrane', symbol $\Delta\mu$; another name for the combination is 'proton motive force', abbreviated as PMF.

During the transduction by chemiosmosis of $\Delta\Psi$ into ATP, protons cross the membrane. Call the $H^+$/ATP ratio number $n$, also named the 'coupling ratio' (Kettner et al., 2003); an extensive literature deals with the coupling ratio, for recent studies see Toei et al. (2007), Tomasek and Brusilow (2000) and Steigmiller et al. (2008). The minimal membrane potential $\Delta\Psi_n$ for ATP synthesis at a ratio $n$ equals

$$\Delta\Psi_n = \Delta\Psi_1 / n.$$

At a potential lower than $\Delta\Psi_n$ the enzyme can only function as a pump that sends protons across the membrane at the cost of ATP (Fig. 6). Assuming the just mentioned value of 360 mV, $\Delta\Psi_3$ becomes 120 mV for $n = 3$, and $\Delta\Psi_5$ becomes 72 mV for $n = 5$. Different proton translocating enzymes have a different value for $n$: the cell uses ATPases with a low $n$ value to acidify cell organelles, for instance for digestion. When a membrane contains two ATP Synthases active with different ratios (in Fig. 7, $n$ = 1 and 2), the result is a net dissipation of ATP (discussed for $Ca^{2+}$ ATPases with different $n$ values (Gafni and Boyer, 1985)). The membrane potential then assumes a value between the equilibrium potentials of the two $n$ values. Such a process has been called a 'futile cycle' (Konings and Booth, 1981). In the V-ATPase of the plant vacuole membrane the coupling ratio can vary from 4 to 2 as the $\Delta\Psi$ increases (Tomasek and Brusilow, 2000). The variation has been attributed to a variable intensity of dissipative proton transport through the ATP Synthase ('slip' or 'intrinsic uncoupling') that would not affect ATP synthesis / hydrolysis (Feniouk et al., 2005). The $F_oF_1$-ATP Synthase in mitochondria and chloroplasts functions at higher ratios, 3 to 5; much higher values ~ 20 have however also been reported (Azzone et al., 1978a, 1978b, 1978c).

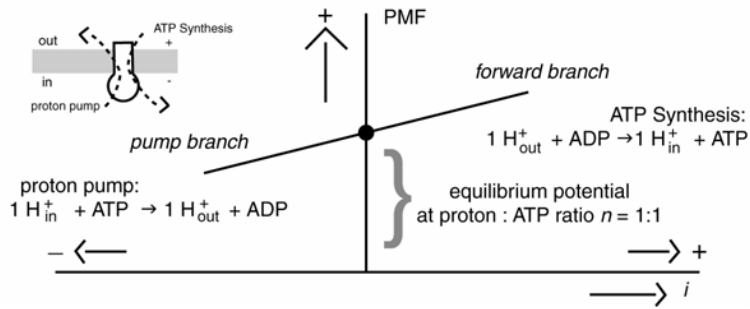

Fig. 6. *I – V curve of a simple ATP Synthase*  The magnitude and direction of the proton current ('I') through ATP Synthase depends on the PMF ('V') and the proton:ATP ratio under which the enzyme operates. At low PMF the enzyme works as a proton pump and consumes ATP, at high PMF it generates ATP.

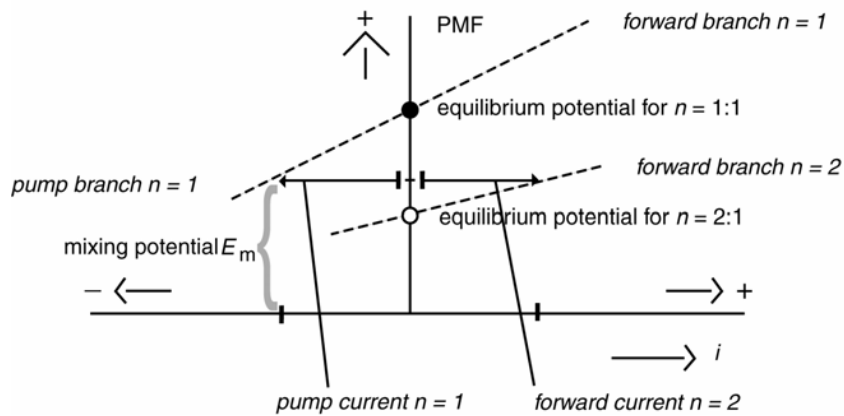

Fig. 7 *Net dissipation of ATP ('futile cycle') when two ATPases with different stochiometries are operating simultaneously, one as ATP Synthase (forwards), the other as ATPase (backwards)*  At the mixing potential $E_m$ the ATP yielded by the forward current equals half the ATP consumed by the pump current.

The ATP Synthase enzyme contains two moieties, the membrane bound $F_o$ part and the water-immersed $F_1$ part (Boyer, 1997). The $F_o$ part contains the translocator of the protons across the membrane. The $F_1$ part is itself composite, containing three $\alpha$ and three $\beta$ subunits that alternate in a ring. The $\alpha$ and $\beta$ subunits are similar, the difference being that whereas the nucleotides on the $\alpha$ subunit cannot be exchanged with the medium, the nucleotides on the $\beta$ subunit can: here the ATP is synthesized (or, conversely, hydrolysed in the reverse process). The ATP synthesis occurs by Boyer's (1979) 'binding change mechanism'. The bound ADP and phosphate spontaneously form ATP in the local dry conditions (DeMeis, 1989). This ATP remains strongly bound. In the energy transducing step, the bound ATP is released when protons move through $F_o$ from one side of the membrane to the other (Fig. 1).

2.3 Thermosynthesis Based on Convection (TSC)

Many heat engines involve thermal transitions, and ATP-generating biological heat engines could plausibly involve thermal transitions in the proteins and the membranes that sustain chemiosmosis. More specifically, the thermal transition would concern the thermal unfolding of the protein ATP Synthase (Muller, 1995) or the thermotropic phase transition of the lipid membrane (Muller, 1993). Thermal cycling would be due to the circulation of the fluid in a convection cell. The previously proposed two heat engines work on this thermal cycling, 'Thermosynthesis by Convection' (TSC). The mechanisms have been applied in models for the origin of life (Muller, 1995, 2005a) and the emergence of photosynthesis (Muller, 1996, 2003, 2005a, 2005b).

During the origin of life, the proposed mechanism of the first protein (FP) is a thermal variation of the binding change mechanism of ATP Synthase (Muller, 1996) that is called 'protein-associated thermosynthesis' (PTS) (Fig. 1). Thermal cycling would have caused the FP to synthesize by 'enzymatic promiscuity' (Black, 1970; Copley, 2003; Kazlauskas, 2005; Khersonsky et al., 2006; Hult and Berglund, 2007) several high-energy condensation products: in addition to ATP and other nucleotide triphosphates, phospholipids, randomly constituted peptides and RNA[1] (Muller 1995, 2005a). The FP would reverse reactions similar to the hydrolyses of amide and P-O bonds supported by the promiscuous enzyme aminopeptidase P (Jao et al., 2004). The contemporary enzyme most closely related to the FP would be the β subunit of ATP Synthase.[2] Today, ATP Synthases still show high (Villaverde et al., 1998; Wang et al., 1993) and low temperature unfolding (Hightower and McCarty, 1996).

A key distinction between the FP and almost all regular enzymes (ATP Synthase is an exception (Vinogradov, 1999)) is that whereas regular enzymes catalyse both the forward and backwards reaction, and diminish the free energy, the proposed ancient FP generated a stable high energy product during thermal cycling. Just as in ATP Synthase today (Vinogradov, 1999), some partial steps of the FP must have been irreversible in order to avoid the decay of the high energy product by the reverse reaction (Muller, 1995), a regulation that would have been controlled by the temperature. This *kinetic irreversibility* should be distinguished from the assumed *conformational reversibility* during thermal cycling of the FP, which contrasts with the irreversibility of thermal denaturation in many of today's proteins. During some partial processes of the thermal cycle the conformational changes would have been irreversible, but at the end of the thermal cycle the original conformation would have been regained.

In the second heat engine mechanism, a change in coupling ratio was synchronized with variation in the dipole potential of the biomembrane during a thermotropic phase transition in the membrane (Muller, 1993). This 'membrane-associated thermosynthesis' (MTS) is a proposed progenitor of bacterial photosynthesis, with as an intermediate link a mechanism based on a light-induced dipole potential effected by light-dark cycling (Muller 1996, 2003, 2005b). An early ATP Synthase would have been able to vary the coupling ratio, for instance with the temperature. By (1) pumping protons across the membrane at a high coupling ratio and low potential (say 5 protons at a cost of 1 ATP molecule), (2) letting a physical process (temperature change, light absorption) increase the membrane potential by evoking a dipole layer with a high enough dipole potential, and (3) letting protons fall down at low coupling ratio and high potential (say 1 ATP per proton), net ATP could be won (in this example, $5 - 1 = 4$ ATPs) (Fig. 8). A variability of the coupling ratio in present day ATP Synthase is consistent with the frequent experimental difficulty of establishing an integer value for it (Tomasek and Brusilow, 2000).

---

[1] Thermosynthesis proposes a temporal order for the emergence of the arms of transfer RNA (Muller, 2005a) that is consistent with the order obtained by phylogenetic structure analysis (Sun and Caetano-Anolles, 2008).

[2] This protein with its long evolutionary history is still of importance: during cancer it is underexpressed (Cuezva et al., 2007).

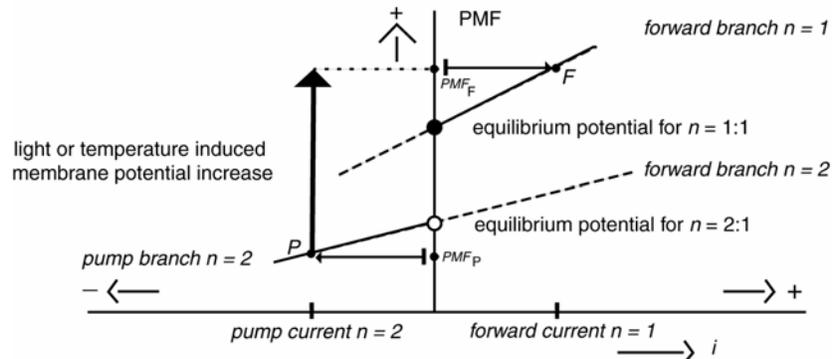

Fig. 8. *A fluctuating membrane PMF (or voltage) permits ATP gain, either from two different ATP Synthases with a different proton-ATP ratio, or from a single ATP Synthase with a variable proton ATP ratio* At high PMF and low proton:ATP ratio, more ATP is synthesized from the forward current than is later consumed by the pump current at low PMF and high proton:ATP ratio when the same charge is pumped again across the membrane. The activity of the concerned enzymes must be properly regulated (switched on and off) by the PMF in order to avoid the dissipation shown in Fig. 7.

An organism carried along by a convective current is passively moved, and photosynthesis may have emerged while organisms were moved in this way. Compared to photosynthesis and respiration, thermosynthesis such as by PTS has a low power, since a single FP yields only one ATP molecule per thermal cycle. The cycle time of a rotation convection cell is expected to be in the range of seconds or longer. The turnover time of FP would be much larger than the turnover time of today's ATP Synthase during photosynthesis or respiration, which can be as low as ~ 10 ms (Cramer and Knaff, 1991). The power (energy produced per unit time) of these last two processes is therefore much higher than the power of thermosynthesis, clearly a selective advantage. On the other hand, the light and food they require are not always and have not always been available at all places and at all times, in particular not, at high latitudes, during the winter near the Poles and, at all latitudes, during the extensive glaciations of the Archean and the Proterozoic.

## 2.4 The Fire: Hydrothermal Vents

After their discovery in 1977 (Corliss et al., 1979), a role of hydrothermal vents in the origin and evolution of life was proposed (Baross and Hoffman, 1985; Corliss et al., 1981). The temperature in black smoker gases leaving the hydrothermal vents can be as high as 350 °C (Corliss et al., 1979); recently (Reed, 2006) a temperature above the critical temperature of seawater, at 298 bar, 407 °C (Bisschoff and Rosenbauer, 1984, 1988), was reported. The at first sight appealing idea that large biomolecules would be formed within the black smokers is however incorrect, since these molecules decompose long before they reach a temperature of 350 °C; even amino acids are not stable at these temperatures (Miller and Bada, 1988).

Due to the thermohaline circulation, the water at the bottom of the ocean is just a few degrees above freezing (Warren, 1981), which results in a large thermal gradient in the boundary layer of the water flowing over the hydrothermal vent. This large thermal gradient decreases with distance to the vent; smaller thermal gradient must be present as well. Moreover, some hydrothermal vents and mounds show maximal temperatures much lower than 350 °C (Kelley et al., 2001; Pirajno and Van Kranendonk, 2005; Reed, 2006). Although temperatures around 100 °C may still be too high to *start* life (Bada and Lazcano, 2002; Miller and Lazcano, 1995):

> RNA and DNA are clearly too unstable to exist in a hot prebiotic environment,

today's organisms can live at such a temperature, and it follows from rRNA sequences that the last common ancestor of all living organisms lived at such a high temperature of ~ 100 °C (Woese, 1987).

Places of lower temperature (~ 80 °C) at hydrothermal vents / mounds have been proposed as sites for prebiotic reactions that generate the substances required for the origin of life (Martin and Russell, 2002; Russell and Arndt, 2005; Russell and Hall, 1997). However, this temperature is still high enough to destabilize nucleic acids. Only a temperature below freezing point is conducive to the formation of stable bases of nucleic acids (Vlassov et al., 2005).

Whereas some researchers favour a hot origin of life (Bahn et al., 2004; Baross and Hoffman, 1985; Fox and Dose, 1977), other origin-of-life researchers favour a cold one (Buford Price, 2007; Trinks et al., 2005). Thermosynthesis suits as bridge for the gap between the seemingly antithetic hot and cold origins since it involves convection, which requires a thermal gradient, i.e. both a high and a low temperature. More generally, the thermosynthesis theory invokes dynamic temperatures as key agent: the origin of life would have required 'thermal turmoil'. The heat produced by episodic meteorite impacts must have made the early Earth's surface temperature far from static.

2.5 The Ice: Glaciations on the Early Earth
In order to understand early glaciations, we consider what determined Earth's surface temperature in the past. Initially it must have been very high due to the heat released during the gravitational collapse that formed the Earth 4,5 Ga. Thereafter, gravitation remained a large heat source during intermittent meteorite impacts; the impact that resulted in the formation of the Moon must have remelted the Earth's surface (Halliday, 2000). The formed Moon was itself hit by meteorites several times around 4,3 Ga, and the Earth cannot have escaped this Late Heavy Bombardment (Chyba, 1993; Cohen et al., 2000; Ryder, 1990). Large impacts sterilized the Earth (Sleep et al., 1989), but ejecta thrown out into space may have contained organisms that returned life to the planet after they fell back[3] during clement conditions (Horneck et al., 2008; Stöffler et al., 2008).

Between large impacts the Earth's surface temperature was the result of a balance between the heat coming in by Solar radiation and the heat going out due to Terrestrial radiation. The latter is diminished when infrared radiation emitted at the Earth's surface is absorbed by greenhouse gases such as methane, $CO_2$ and water vapour (Khalil, 1999).

The early Sun was ~30% less bright than today (Tajika, 2003) due to a lower internal temperature caused by a higher hydrogen/helium ratio (Gough, 1981). The early Earth may therefore have been cold and covered with ice (Bada et al. 1994; Sagan and Mullen, 1972; Valley et al. 2002). Again, frequent meteorite impacts may have melted this ice (Bada et al. 1994). On the other hand the early Earth may have been kept warm by the greenhouse gas methane (Kharecha et al., 2005) produced by early methanogens that gained free energy from the reaction (Thauer, 1998)

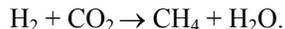
$$H_2 + CO_2 \rightarrow CH_4 + H_2O.$$

The evidence available in the geological record diminishes as one goes back in time, and it seems hard to decide on a hot or a cold ocean before 3 Ga (Kasting and Howard, 2006; Kasting and Ono, 2006). Early global glaciations have been dated at ~2,9 Ga (Young et al. 1998) and ~2,2 Ga (Evans et al., 1997) and may have been due to a loss of methane. Because of the constant loss of hydrogen to interplanetary space the Earth's atmosphere must steadily have become less reducing (Kasting et al., 1993). Oxygen may have been generated by cyanobacteria that possibly emerged at 2,8 Ga (Schwartzman et al., 2008), i.e. after the 2,9 Ga glaciation. The oxygen concentration in the atmosphere became significant between 2,45 and 2,32 Ga (Bekker et al., 2004). The first fossils attributed to the eukaryotes have been dated around 1,8 Ga (Knoll et al., 2006), after the 2,2 Ga glaciation, and it is proposed that their progenitors lived during the glaciations on thermosynthesis by the mechanisms described hereafter.

---

[3] Ancient organisms 'put on ice' in meteorites of terrestrial origin, either in orbit around the Sun or deposited in repositories on the Moon (Armstrong et al. 2002) (especial near the poles in crater interiors shaded from the Sun) or other celestial objects, could allow the testing of the models presented here. This refutes the objection against these models that they are in principle untestable.

The dark bottom of the ocean may have remained reducing long after 2,2 Ga (Huston and Logan, 2004). In that case there would often have been no means of primary production by photosynthesis and chemosynthesis in organisms on top of hydrothermal vents at the bottom of the ocean, and no alternative to thermosynthesis as primary producer. It is often claimed that the thermal gradient of a hydrothermal vent is a potential biological energy source and that it in particular may have driven the origin of life (Baross and Hoffman, 1985; Corliss et al., 1981):

> . . . heat derived from the hydrothermal cooling of newly-emplaced oceanic crust provides the energy source for converting inorganic precursors into organic compounds.

The gradient may similarly have been a biological energy source during global glaciations. *Such biological use of heat as an energy source requires specific biochemical pathways and physiological adaptations, which remain to be identified.* The world around us should be explored for these pathways and adaptations that may still be extant but would have remained concealed. They may however also have gone extinct, in which case they would require theoretical and conceptual development.

In addition to the biological heat engines proposed by this author, few other biological heat engines have been proposed. Sidney Fox drew attention to the role of heat as an energy source during the synthesis of proteinoids from amino acids (Fox and Dose, 1977), but nowadays most origin of life investigators seem to consider his approach a dead end in origin of life research; for an exception, see Bahn et al. (2004). A role of heat during the unfolding of ATP Synthase that results in ATP release by the binding change mechanism (Vekshin, 1991) and during photosynthesis (Komissarov, 1990) has been proposed by other authors, who did not recognise that the underlying mechanism must concern a heat engine. Matsuno has proposed several heat engine mechanisms (1995, 1996) and has experimentally investigated quenching of a high-temperature chemical equilibrium in a simulation of a hydrothermal vent (Imai et al., 2005).

In this study and those following, we consider in detail heat engine mechanisms based on the chemiosmotic machinery and apply them in models that permit life to survive and evolve at hydrothermal vents in the absence of photosynthesis and respiration.

## 3. Between Fire and Ice: Thermosynthesis in a Thermal Gradient (TSG)

3.1 Thermal Fluctuations on the Surface of a Hydrothermal Vent
Since convection above it is plausible, a hydrothermal vent was a suitable niche for early thermosynthesizers that lived on TSC. Convection is self-organizing, and arises in a fluid subject to a temperature gradient when the upward force due to buoyancy—in turn due to the density difference caused by the gradient—overcomes the friction caused by the viscosity of the fluid. Being carried along by convection does not cost an organism energy.

Geysers such as Yellowstone's well known 'Old Faithful' demonstrate the periodicity of physical processes in volcanic areas; they are good examples of relaxation oscillations (Van der Pol and Van der Mark, 1928). Near a hydrothermal vent thermal cycles with large amplitudes can be expected due to a similar pulsation in the ejected fluids (Kompanichenko et al., 2008). An additional agent of temperature variation could be turbulence in the water flowing over the vent.

The proposed scenario for the emergence of an organism with a thermotether starts with a sessile thermosynthesizer that benefited from these temperature fluctuations upon attachment to the hydrothermal vent. Thermosynthesis in a thermal gradient (TSG) is characterized by ATP generation *without* the organism being moved by convection.

3.2 Emergence of the Thermotether: a simple Relaxation Oscillator

A fluid flow can be characterized by the Reynolds number *Re*, the ratio of inertial and viscous (frictional) forces:

$$Re = v\,L\,\rho\,/\,\eta = v\,L\,/\,\upsilon\;,$$

where *v* is the speed of the fluid, *L* a relevant length, $\rho$ the fluid density, $\eta$ the fluid viscosity, and $\upsilon = \eta\,/\,\rho$ the dynamic viscosity of the fluid (Purcell, 1977).

The flow adjacent to the surface of a rigid object is laminar, whereas it becomes turbulent in the bulk of the fluid. The area of laminar flow is called the boundary layer, and when the object and the fluid are at different temperatures, this boundary layer comprises most of the temperature difference. The boundary layer at the interface between the ocean and a hydrothermal vent is expected to contain a similar thermal gradient.

The edge between the boundary layer and the bulk fluid is not sharp, which makes the thickness $\delta$ of the boundary layer a somewhat loosely defined notion. In the case of a plate parallel to the flow the relation $\delta \approx x\,/\,\sqrt{Re}$ has been given, where *x* is the distance to the edge of the plate, and *Re* the Reynolds number (Fig. 9) (Richardson, 1989). The boundary layer therefore starts thin. For Reynolds numbers of $10^4 - 10^6$, the boundary layer thickness at 1 cm from the edge of the plate would consequently be $10^{-4} - 10^{-5}$ m, i.e. 100 – 10 µm.

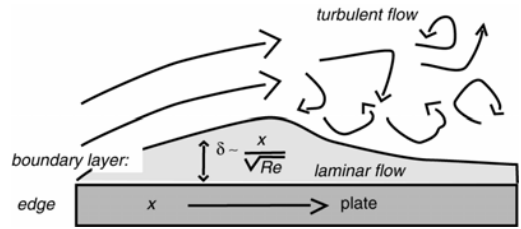

Fig. 9. *Boundary layer above a plate in a flowing fluid* The flow is laminar within a distance $\delta$ from the plate; $\delta$ increases with the distance *x* to the edge of the plate.

*Proposed mechanism of the thermotether* Because active movement in a viscous fluid is dissipative, it would cost an organism of the size of say 1 µm much energy to move inside the boundary layer (Purcell, 1977). For a smaller organism, movement becomes even more difficult because of the lower value of the Reynolds number: below the scale of bacteria all movement is strongly overdamped (Howard, 2001), and instigating active biological movement on this scale seems impossible. But in general, biological phenomena start small— either during the life cycle of the organism or during speciation—and then increase in size. *How did active biological movement emerge when it could not start small?* This study proposes: *Active biological movement emerged upon the 'kinetic takeover' by life of movement that resulted from thermal conformation changes driven by relaxation oscillations in the thermal gradient above a hydrothermal vent.*

In the proposed scenario the vent-attached microorganism acquired a flexible tether (Friedel et al., 2006) that permitted the microorganism to move at random in the medium near the vent. This random movement enhanced the amplitude of the experienced thermal cycling, which increased the free energy generated by thermosynthesis.

In the next evolutionary step the flexible tether turned into a self-actualized protein engine that cyclically changed conformation in the thermal gradient by making use of cold denaturation, the unfolding of a protein at low temperature around the freezing point of water (Brandts, 1967).

When the so defined 'thermotether' extended into the cold beyond the boundary layer it contracted due to cold denaturation. After shortening, it found itself near the hot substrate, and extended again. At no cost of free energy for the organism, this relaxation oscillation thermally cycled the FPs that the organism contained, thus enabling these to generate ATP. The thermotether, a long (~ µm scale) filamentous protein, had taken over the role of convection. This relaxation oscillation requires a delayed adaption by the protein to the equilibrium conformation at the local temperature; the delay may be due to the time required for diffusion in conformational space (Kramer, 1940) and may also be associated to co-movement of water with the protein (Frauenfelder et al., 2006).

The capability of making use of a static thermal gradient opened up areas at a larger distance from the vent as habitat, where fluctuating temperatures were absent. It would have been advantageous when the length of the thermotether increased so that it spanned the entire thermal gradient across the boundary layer. Because of the linear increase of the boundary layer with the distance to the edge (Fig. 9) the thermotether could gradually increase in size and the organism its distance to the edge (Doebeli and Dieckmann, 2003) during evolution.

The presence of a thermotether supposes export of synthesized proteins through the cell membrane. The involved translocation capability must be old. It has been proposed that a first ribosome worked on thermal cycling (Muller 1995, 2005a), and this may also have been true for the first protein exporter. Export makes use of translocon protein complexes (White and VonHeijne, 2008). Hereafter protein export will be associated with the type III secretion machine; the related hollow bacterial flagellum can internally transport monomers of its constituting protein to its tip (Cornelis, 2006; Galan and Wolf-Watz, 2006; Holland, 2004; MacNab, 2004; Wickner and Schekman, 2005).

*Implications* There have been few studies of heat engines consisting of protein. It has been proposed that a macroscopic engine constructed of collagen that operated on a salt concentration difference also would work on a temperature difference (Steinberg et al., 1966).

Any spatially extended biological structure could have functioned as a thermotether, even an entire organism could. Filamentous structures that may be microfossils are ubiquitous and date back as far as 3,5 Ga (Hofmann et al., 2008). Moreover, (Reysenbach and Cady, 2001):

> . . . all of the microfossils associated with ancient hydrothermal deposits reported
> to date are filamentous.

These filamentous fossils resemble today's prokaryotic cyanobacteria, but it has been recognized that this morphology does not imply a photosynthetic physiology (Schopf, 2006). Fossils of long filaments ( 0,5—2 μm wide, up to 300 μm long) are also found near the 3,235 Ga old remains of hydrothermal vents (Rasmussen, 2000). The filaments may even have been caught frozen while functioning as thermotether (Rasmussen, 2000):

> Perhaps most significantly, the changes in preferred orientation of filaments in
> different areas . . . may indicate behavioural variations in different micro-
> environments, a distinctly biological feature. Similar patterns are evident in
> modern microbial mats . . . and hot-spring communities . . . indicating a biotic
> tropism or taxis in response to a gradient or fluctuation in a salient ecological
> parameter.

Today's hydrothermal vents could be explored for the presence of organisms with thermotethers, and the ubiquitous filamentous organisms especially (Hofmann et al., 2008) investigated for TSG capability. Organisms can live on the oxidation of basalt (Santelli et al., 2008) and respirers may easily win the competition with thermosynthesizers where the water of the ocean is oxygenated. Such competition may be absent in those hydrothermal vents where the natural water is anoxic, such as intermediate depths of large parts of the ocean, the Pacific coasts of the Americas, of Namibia and in the Black Sea, Baltic Sea and the Arabian Sea (Levin, 2002). Here large filamentous bacteria are common. In many of these places organisms that respire will use oxidants different from oxygen, such as sulphate, but in some places a redox equilibrium that would impede all respiration is plausible that would give room for thermosynthesizers. If present day TSG organisms could be found, their isotopic footprints such as their $^{13}C/^{12}C$ ratio should be determined. An isotopic footprint distinct from photosynthesis could link TSG to the oldest fossils, which show very low $^{13}C/^{12}C$ values (~ −30‰) (Horita, 2005).

Life started small. During TSC there is no evolutionary driving force towards larger size. On the contrary, being small has the advantage of a fast heat exchange between the organism and its local environment. The small size moreover means that no internal transport mechanisms are needed (Schulz and Jørgensen, 2001). In contrast, in TSG a large size has the advantage that a larger part of a thermal gradient is spanned. In high-nutrient media, where there is no diffusion limitation, large bacteria exist. In the gut of tropical fish bacteria as large as $80 \times 600$ μm are reported (Schulz and Jørgensen, 2001). Today, hydrothermal vents (Jannasch and Wirsen, 1981; Kalanetra et al., 2004; Nelson et al., 1989; Schulz and Jørgensen, 2001) and low-oxygen areas of the ocean (Levin, 2002) contain some of the largest bacteria known.

The spasmoneme of *Vorticella* rotates during a $Ca^{2+}$ -induced contraction (Moriyama et al., 1998). Such an effect of $Ca^{2+}$ can be interpreted as a lowering of a thermal transition temperature, i.e. as a way to mimic an increase in temperature while the temperature remains constant (Muller, 1996). The relationship between temperature-induced contraction and rotation suggests the possibility of modelling the emergence of the rotating flagellar motor in terms of the thermotether as well.

## 4. The Flagellar Motor, The Flagellar Pump, The Flagellar Ratchet and the Flagellar Computer

Heat dissipation by heat conduction in a thermal gradient is intrinsic to the thermotether. It is therefore, just as thermoelectricity, thermal diffusion and the thermal diffusion potential, a phenomenon from non-equilibrium thermodynamics (De Groot and Mazur, 1962; Haase, 1963). Requiring only two components, the thermotether and the β subunit of ATP Synthase, the free energy generating machinery is simple, but the energy conversion efficiency—the ratio of the free energy produced by TSG and the heat transported from high to low temperature— can only be very low. In an early thermotether-based organism there was much room for improvement towards higher efficiency, larger complexity and use of smaller thermal gradients.

Movement by protein conformational changes driven by thermal cycling is not subject to impediment by a low Reynolds number. In contrast, it is hard to account for the overdamping of motion on the molecular scale in models that do not make use of thermal cycling (Howard, 2001). In our model, the thermally induced thermotether folding/unfolding would occur concomitantly (Baker, 2004) with binding/unbinding of nucleotides in the subunits of the $F_1$ moiety during reversible ATP hydrolysis and synthesis. Although the thermotether as a whole must function irreversibly, some partial processes can have approached reversibility.

Previous studies (Muller 1995, 2005a) demonstrated that invoking thermal cycling opens up new possibilities in the modelling of early evolution. We now consider additional constructs derived from the thermotether: a second model for ATP Synthase, and models for the emergence of the 'flagellar pump', the flagellar motor, and the 'flagellar ratchet', a biological analogue to Feynman's ratchet mechanism (Feynman et al., 1963). The information processing capabilities of a bacterium can be related to the flagellar motor, and the obtained 'flagellar computer' is compared to the Turing machine (Turing, 1937).

4.1 Emergence of ATP Synthase in a Thermal Gradient

A previous study has given a scenario for the emergence of an ATP Synthase with a temperature-dependent $H^+$/ATP ratio: figures 7 and 8 in Muller (1995). In the scenario a proton translocator is added to a progenitor of the $\alpha$ subunit in $F_1$, and this translocator is pushed/pulled across the membrane during conformational transitions in the $\alpha$ subunit between strong and loose ATP binding (Fig. 10). A variation in membrane thickness (Andersen and Koeppe, 2007) with the temperature, plausible during a thermotropic phase transition in the membrane, occludes a proton binding site of the translocator during part of the thermal cycle. This occlusion leads to the pumping of protons, and the resulting proton gradient permits the import of nutrients. The addition to the proton pump of the $\beta$ subunit that could exchange nucleotides yields an ATP Synthase with a temperature-dependent variable $H^+$/ATP ratio as pictured in Fig. 8. Figure 3B shows how the combination of a proton pump and ATP Synthase generates ATP.

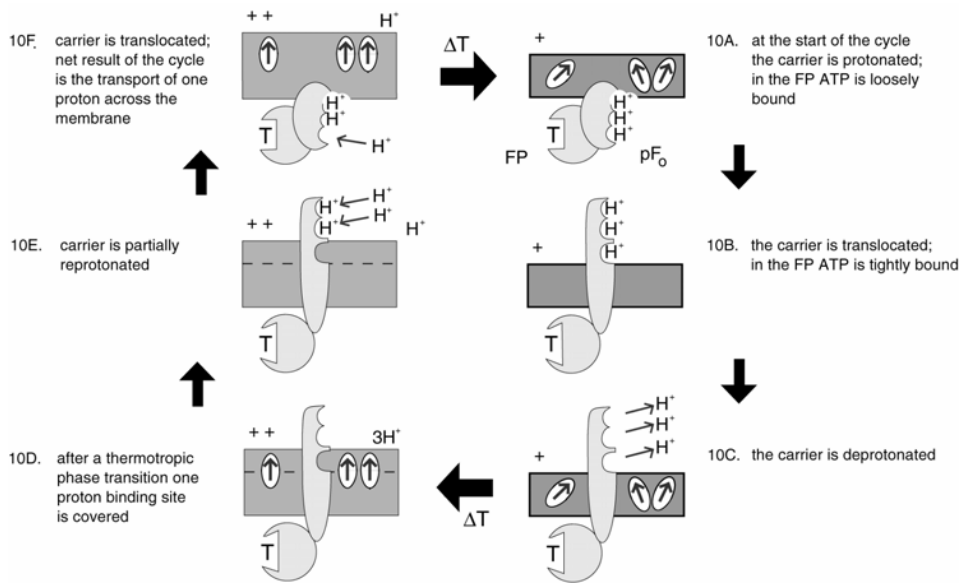

Fig. 10. *Previously proposed mechanism for a proton pump that worked on thermal cycling* During a thermal cycle a thermotropic phase transition caused variation in the thickness of a biomembrane, which occluded a proton binding site in a protein that was translocated across the membrane. The translocation was driven by conformational changes in an FP between strong and loose nucleotide binding. From Muller (1995).

Near the surface of *high*-temperature hydrothermal vents the pH is *low*, near *low*-temperature vents the pH is *high* (Russell, 2007). From a steep pH gradient free energy can be gained in a simple way. An ATP Synthase-containing vesicle that was cycled between two different pHs could generate ATP by what we call the 'Jagendorf mechanism': after the internal pH has equilibrated with the external pH, a decrease in external pH after a movement driven by the thermotether results in ATP synthesis (Jagendorf and Uribe, 1966) (Fig. 11).

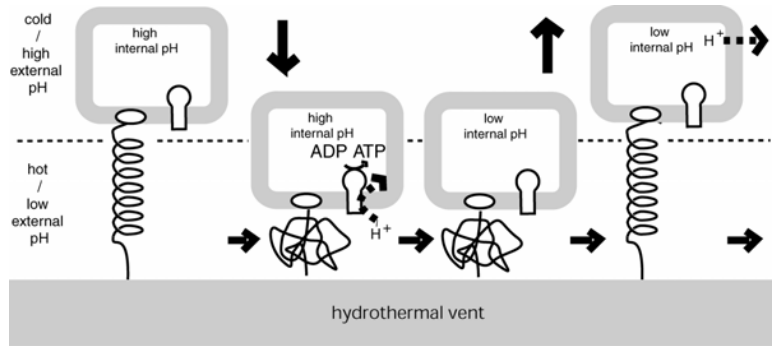

Fig. 11. *Proposed Jagendorf mechanism* Near a hydrothermal vent a pH gradient is present. During a thermotether driven oscillation the external pH of a cell will therefore vary. ATP could be gained from this pH variation, when ATP Synthase is active only at high temperature, with the cell near the vent: ATP is synthesized while protons enter the cell. These protons leave again later by conduction through the cell membrane, when the cell is at its largest distance from the vent, where the external pH is high.

The pH gradient may also have played a role during the *emergence* of the thermotether and ATP Synthase. The protein unfolding temperature typically depends on the pH (Simpson and Finney, 1930), and a thermotether can gradually have gained the ability to oscillate in a combination of a thermal and pH gradient, and eventually to oscillate in a pH gradient only.

The second scenario for the emergence of a proton pump with the same components as ATP Synthase is based on the recently found aqueous proton channels through ATP Synthase's subunit *a*; these lead to the proton carrying Asp-61 residue in the *c* subunit of $F_o$ (Angevine et al., 2003), which can undergo a conformation change (Vincent et al., 2007; Vorburger et al., 2008). The $F_o$ moiety of ATP Synthase contains 10 - 15 of these *c* subunits (Vincent et al., 2007). In the second scenario the conformation change drives a proton pump (Fig. 12), just as in the first scenario described in Fig. 10. No use is made of membrane thickening: the key step is performed by proteins instead of lipids.

A significant pH difference across a membrane is common, and is concentrated across it, as protons freely diffuse in water but not through lipids.[4] A *c* subunit that oscillates on a pH gradient across a biomembrane (Fig. 12) could have evolved from a thermotether adapted to oscillating in this pH gradient. This mechanism may still play a role in today's ATP Synthase: the strain generated in the *c* subunit would release the tightly-bound ATP in the $F_1$ moiety, constituting the conformational energy transduced from $F_o$ to $F_1$ depicted in Fig. 1A.

It is proposed that the flagellar pump emerged when similar proton-transferring proteins attached to a protein export apparatus, and caused this apparatus to pump protons and later to rotate, or, in a slight twist of this idea, that the protein export apparatus itself extended and started to function as a thermotether and later as flagellum.

---

[4] Biological use of a thermal gradient *across* a biomembrane has been proposed (Gaeta et al., 1987), but is considered less plausible. The temperature difference across a biomembrane can only be small, given its thickness of a few nm. As the difference between the *thermal* conductivities of a lipid membrane and water are expected to be small, a temperature gradient will not become concentrated in the membrane; in contrast, the high *electrical* conductivity of water does cause electric fields to be concentrated across the membrane.

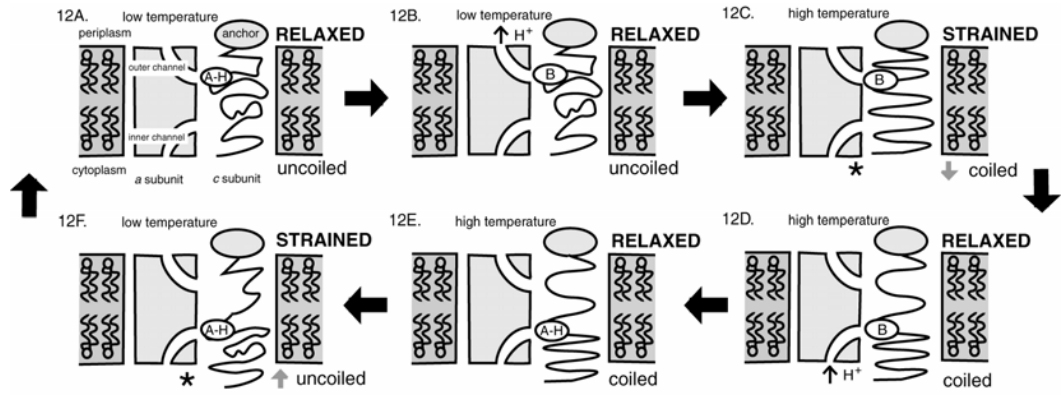

Fig. 12. *A proton pump based on the thermotether* Instead of using a variation in biomembrane thickness as in Fig. 10, a conformational change is invoked in a proton-translocating protein resembling the thermotether. This protein, which also resembles the *c* subunit in ATP Synthase, was anchored in the periplasm.

12A → 12B. At low temperature, the acid residue A-H loses its proton to the periplasm. The proton has been transported. The base residue B stays behind.

12B → 12C. The temperature increases. The cold-denatured protein folds, but it cannot instantaneously assume its equilibrium shape — its (compression) strain is indicated by the asterisk.

12C → 12D. While the protein extends, the base residue B is positioned in front of the inner proton channel leading to the cytoplasmic space.

12D → 12E. At high temperature the base residue B is protonated along the proton channel from the cytoplasm: the acid residue A-H is formed.

12E → 12F. The temperature decreases. Unfolding at low temperature yields a strained protein in tension.

12F → 12G. The cycle closes when the protein relaxes to its cold-denatured equilibrium state. The acid residue A-H moves in front of the proton channel leading to the periplasm side, and the cycle is closed.

## 4.2 Today's Flagellar Motor

Before we consider the emergence of the flagellar motor (Silverman and Simon, 1974), we sketch its structure (Fig. 13) and function. It consists of many proteins and is still the subject of extensive research, since '. . . we do not know how the motor actually works . . . ' (Berg, 2003). Although its present day functioning has not yet been completely resolved, we assume that its emergence can already be modeled.

The motor is important. In *Escherichia coli* the flagellin protein FliC of the rotating helical filament comprises 8 % of the bacteria's total protein (Neidhart, 1987). During the growth of the flagellum, monomers of FliC move through the hollow flagellum to its tip by an ATP dependent process (Chng and Kitao, 2008; MacNab, 2003). This movement is similar to protein export and we already referred to the large resemblance of the core of the flagellar motor to the type III protein export apparatus (Cornelis, 2006; Holland, 2004; MacNab, 2004; White and VonHeijne, 2008; Wickner and Schekman, 2005). The FliI component of the flagellar motor (Imada et al., 2007) is homologous to the α and β subunits of ATP Synthase (Berg, 2003).

The rotation rate is several hundred Hz (Berg, 2003), about the same turnover frequency as ATP Synthase. Effecting movement does not require a particular shape of the flagellum. At low Reynolds numbers, anything rotating will do (Purcell, 1977):

> The interesting question is not how they swim. Turn anything – if it isn't perfectly symmetrical, you'll swim.

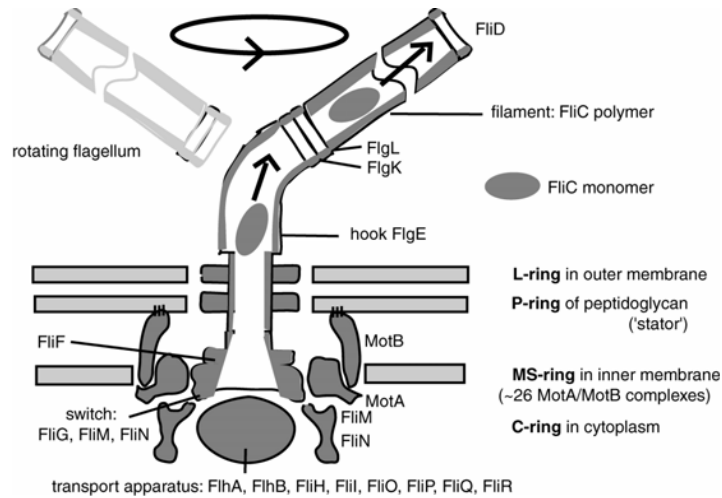

Fig. 13. *Today's flagellar motor* (E. coli*) and some of its proteins*  The rotating flagellum has a length of ~1 μm. The flagellum consists of a hook and a filament of FliC monomers; growth occurs by addition of FliC monomers that have moved through the hollow flagellum to the tip. The transport apparatus sends the FliC monomer across the cytoplasmic membrane.

EM pictures show several rings in the cell wall. The P-ring in the peptidoglycan is associated with the stator of the motor and is connected to the MotA/MotB protein complex. The MotB protein contains the residue involved in proton translocation and is part of the MS-ring in the inner membrane. The C-ring consists of the FliM and FliN proteins. The FliG-FliM-FliN complex implements inversion of rotation direction.

A key question in the emergence of the motor is therefore the origin of torque, which is associated with the proton-translocating MotA/MotB proteins that are anchored to the cell wall. FliG couples the MotA/MotB proteins on the stator to the rotor. Whereas bioenergetic processes as a rule are driven by ATP, the motor is driven by the PMF, just as ATP Synthase (Larsen et al., 1974 ). It has already been mentioned that when the PMF consists of a ΔpH, ATP synthesis by ATP Synthase is endergonic (Hardt, 1980), and this is true for the flagellar motor as well (MacNab, 1979).

Another important question is the mechanism of the control of the directed movement. During chemotaxis the bacterium 'runs', i.e. swims steadily in one direction; then it 'tumbles', performs rotation diffusion, and then 'runs' again in another direction. When the new direction is advantageous, the next 'tumble' is delayed, otherwise it is advanced (Webre et al., 2003). Flagella in *E. coli* are bundled and move as a whole. Tumbling occurs when a single flagellum upsets the bundle by temporarily switching between rotating clockwise (CW) and counter clockwise (CCW), 'as viewed along a filament from its distal end toward the cell body' (Berg, 2003). The 'switch complex' consists of FliG, FliM and FliN (Brown et al., 2007; Cohen-Ben-Lulu et al., 2008; Mashimo et al, 2007).

### 4.3 The Flagellar Pump

*Proposed emergence*  Microorganisms are easily tethered to substrates by their flagella (Silverman and Simon, 1974). In the proposed model for the emergence of the flagellar motor, its progenitor was a flagellar proton pump that worked on thermal cycling (Fig. 3C). The protons generated ATP when they returned through an ATP Synthase.  Proton translocation may have been assisted by torque generated in the thermotether during thermal cycling; rotation in one direction would have been obtained by connecting a ratchet and pawl, or in short, a ratchet (Fig. 14) to the flagellum that functioned as an axis.

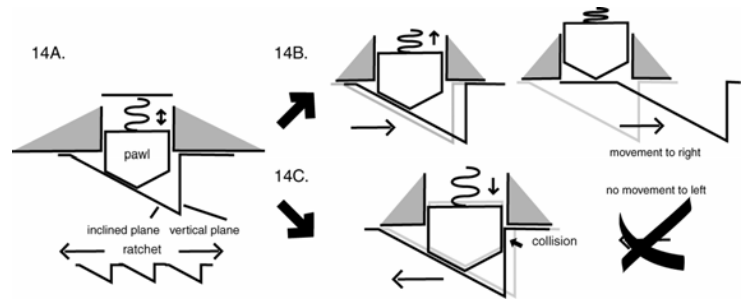

Fig. 14. *Asymmetry in a ratchet-and-pawl mechanism* Sliding of one surface over another is permitted in one direction only. A pawl sticking out from inside the top surface falls into an asymmetric notch in the bottom surface. The inclined plane inside the notch can push the pawl back when the lower surface moves to the right with respect to the upper surface (14B). When the surfaces moves in the opposite direction, the vertical plane inside the notch catches the pawl, and halts further movement to the left (14C).

The first flagellar pump may have contained parts taken from ATP Synthase when this acted as proton pump: multiple proton-donating units (*a* subunit), and multiple proton translocators (*c* subunit) would have been attached to the inner surface of the ring that functioned as a stator and evolved into today's MotA/MotB protein (Fig. 15, 16) with a Asp-32 residue that carries the translocated proton. The MotA/MotB protein combination indeed shows conformational changes (Kojima and Blair, 2001; Schmitt, 2003) during proton translocation and could have applied torque to the type III protein export apparatus (MacNab, 2004) that constituted the rotor.

Thermotaxis is common in bacteria; some move towards a heated wire (Metzner, 1920). *E. coli* aggregates in a thermal gradient (Paster and Ryu, 2008). It should be observable whether microorganisms that show thermotaxis towards a heated wire attach their flagellum to it and pump protons while the flagellum oscillates in the thermal gradient.

4.4 Switch to Isothermy: Emergence of the Flagellar Motor
The model for the emergence of the flagellar motor starts with a dissipative system: while heat moves from high to low temperature in a thermal gradient, oscillations driven by the thermal gradient in protein subunits enable a flagellar pump to function. During a glaciation, evolution maximized the efficiency of the flagellar pump, which initially was very low but eventually reached reversibility, at least for some partial steps of the mechanism—the pumping mechanism being an intrinsically dissipative process, overall reversibility was out of reach.

    The return of reductants and oxidants generated by photosynthesis after the end of the glaciation again enabled regular proton pumping by respiration. It could have been effected by endosymbiosis with a respirer or by the formation of a chimera involving a respirer. Proton pumping by the flagellar pump then became redundant; instead, the newly available protons caused the pump to reverse and to co-adapt it as the flagellar motor. Upon disconnection of the tether from its substrate, it started to function as a flagellum while contracting and rotating (Fig. 3D, Fig. 17).

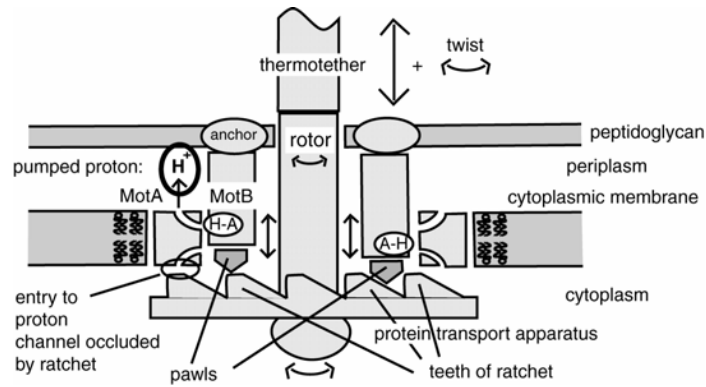

Fig. 15. *Overview of the proposed flagellar pump, a progenitor of the flagellar motor* During the thermal cycling sustained by the thermotether the MotB protein takes up protons from the cytoplasmic side of the membrane and sends them across the membrane (details are given in Fig. 16). Thermal cycling also may have moved back-and-forth or rotated a ratchet that assisted in the translocation of the MotB protein. It moreover may have occluded the entry to the inner proton channel in the MotA protein.

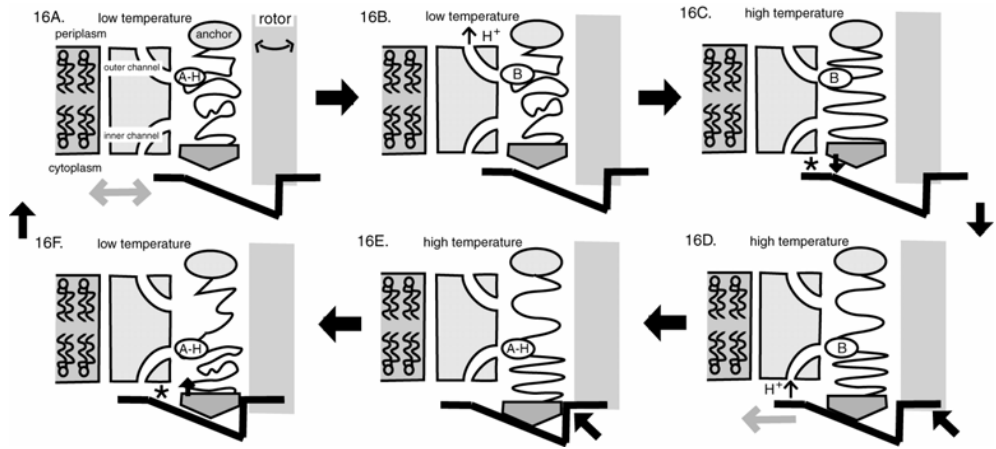

Fig. 16. *Cycle of the flagellar pump* Proton translocation involves the MotB progenitor protein. The mechanism resembles the mechanism of the proton pump described in Fig. 12. The MotB protein is anchored to the peptidoglycan membrane in the periplasm and contains a moving 'pawl' in the cytoplasm.

16A → 16B. At low temperature, the acid residue A-H loses its proton to the periplasm, the base residue B remains.

16B → 16C. The temperature increases. The cold-denatured MotB folds and assumes a strained (asterisk) compressed configuration.

16C → 16D. MotB lengthens. The base residue B moves in front of the inner proton channel that leads to the cytoplasm. The pawl hits the sliding plane of the notch in the ratchet, pushing the ratchet to the left.

16D → 16E. Still at high temperature the base residue B of the extended and folded protein is protonated along the proton channel from the cytoplasm side, forming the acid residue A-H. The pawl pushes the ratchet so far to the left that it occludes the proton channel. The ratchet halts when the pawl hits the vertical plane inside the notch.

16E → 16F. The temperature decreases. Unfolding at low temperature yields a strained MotB protein in tension that contracts.

16F → 16A. The cycle closes when the MotB protein relaxes to its cold-denatured equilibrium state. The residue with the base residue B is positioned in front of the proton channel leading to the periplasm. The ratchet has lost contact with the pawl and can diffuse freely.

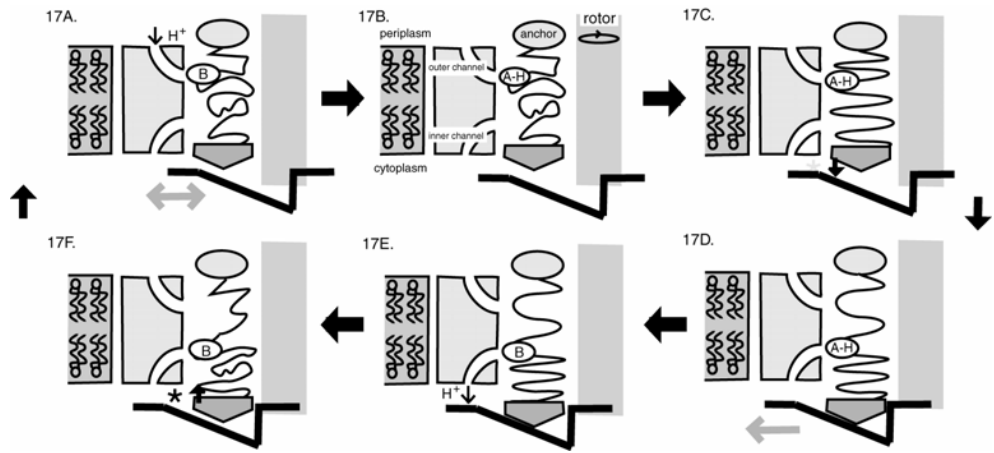

Fig. 17. *Cycle of the flagellar motor* The cycle is the reverse of the cycle of the flagellar pump. In addition, the MotB protein undergoes conformational change by protonation instead of by a change in temperature.

17A → 17B. The base residue B takes up a proton from the periplasm and forms the acid residue A-H.

17B → 17C. The MotB protein folds and assumes a strained, compacted conformational state, indicated by the asterisk.

17C → 17D. MotB expands. The acid residue moves in front of the inner channel. The pawl makes contact with the sliding plane in a notch of the ratchet, forcing the ratchet to the left until the pawl collides with the vertical plane in the notch of the ratchet.

17D → 17E. The acid residue A-H of the extended and folded protein deprotonates along the inner proton channel.

17E → 17F. MotB unfolds, first to a strained conformation. The base residue B is positioned in front of the outer proton channel. The pawl loses contact with the sliding plane of the notch in the ratchet. The ratchet can diffuse freely.

17F → 17A. The cycle closes when MotB relaxes to its cold-denatured, unfolded equilibrium state. The residue with the base residue B is positioned in front of the proton channel leading to the periplasm.

Sequence data show that (Liu and Ochman, 2007):

> core components of the bacterial flagellum originated through the successive duplication and modification of a few, or perhaps, even a single, precursor gene.

and

> The earliest proteins are proximate to the cytoplasmic membrane with later proteins situated distally, first spanning the outer membrane and then giving rise to structures (i.e., the hook, junction, filament, and capping proteins) that extend outside of the bacterial cell. Thus, the flagellum represents a case whereby its order of assembly recapitulates its evolutionary history.

At low temperature CW is the preferred state, at high temperature CCW (Hasegawa et al., 1982; Turner et al., 1996). This study interprets this temperature effect as a possible relic of the flagellar pump. Its flagellum/thermotether rotated CCW upon contraction after a temperature decrease and CW upon expansion after a temperature increase. When the flagellar pump reversed, and started to function as a flagellar motor driven by the PMF, this would have caused the flagellum to change orientation as well, i.e. to rotate CW at low temperature and CCW at high temperature.

4.5 The Flagellar Ratchet and Feynman's Ratchet

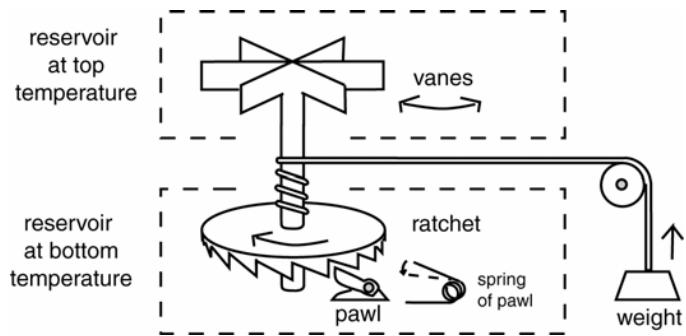

Fig. 18. *Feynman's ratchet* The device invokes a ratchet-and-pawl combination in a thought experiment that tries to circumvent the Second Law of thermodynamics (Feynman et al. 1963). A set of vanes undergoing fluctuating forces due to thermal movements of a gas is connected by an axle to a ratchet-and-pawl. Since the ratchet would only be able to turn in one direction, the axle would turn in one direction as well and lift the shown weight at no cost of free energy. Further analysis shows however that the pawl has a finite chance to jump, which would cause the ratchet to fall back. The Second Law remains valid. When the temperature of the vanes is higher than the temperature of the ratchet, the device can however perform external work and lift the weight. It then works as a heat engine. Feynman has also pointed out that the vanes are set in motion when the vanes have a lower temperature than the ratchet-and-pawl. The hot pawl frequently jumps over the barrier during fluctuations and then forces the ratchet to rotate in a direction opposite to its regular rotation.

We compare the flagellar pump with Feynman's ratchet mechanism for a primitive heat engine (Feynman et al. 1963) (Fig. 18). Both the flagellar pump and the Feynman ratchet (1) generate free energy from a temperature difference and contain (2) a structure interacting with the medium (the flagellum resp. the vane), (3) an axle, and (4) a rotor and stator. One difference is that in the pump the axle cyclically contracts, whereas in Feynman's ratchet the axle is rigid. Another difference is that the MotA/MotB proteins in the stator of the flagellar pump are thermally cycled, whereas the pawls in the stator of Feynman's ratchet undergo isothermal fluctuations.

The flagellar pump including its host, the bacterium, is thermally cycled as a whole. In contrast, a thermal gradient is present across the 'flagellar ratchet', a theoretical device that is an analogue to Feynman's ratchet. Thermal cycling occurs in only those proteins where it matters. The modifications required for the flagellar pump are:
1. the flagellum must have a higher temperature than the stator/rotor;
2. the thermal fluctuations in the torsion of the flagellum must not become excessively damped during their transfer along the flagellum to the rotor/stator (note that energy transfer by changes in torsion is a plausible means of conformational energy transfer between the $F_o$ and $F_1$ parts of ATP Synthase as well);
3. the MotA/MotB 'pawl' must undergo significant entropy-decreasing fluctuations in its conformation. Such entropy-decreasing fluctuations have been observed in small macromolecules (Wang et al., 2002; Ritort, 2003).

The flagellar ratchet has several interesting properties. It resembles the recently proposed theoretical 'Brownian gyrator' (Filliger and Reimann, 2007) that rotates in a thermal gradient. Moreover, a sessile organism with a flagellar ratchet could gain free energy not only when the flagellum has a higher temperature than the stator, but also when the flagellum is subject to non-thermal fluctuations or 'noise' (Fig. 19). It might even function as a turbine in the presence of a constant current (Fig. 3G),[5] a phenomenon that will not occur under the objective of a microscope, because (just as a thermal gradient) the required current is absent there.

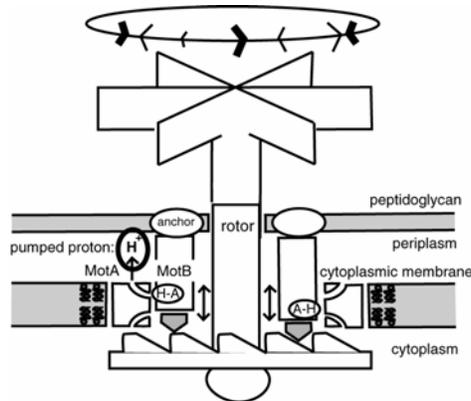

Fig. 19. *The flagellar ratchet* Feynman's ratchet-and-pawl mechanism can function as a heat engine (Fig. 18). It is proposed that when the flagellum of a flagellar motor has a higher temperature than the rotor/stator combination, it can similarly function as a heat engine, which is called a 'flagellar ratchet'. The free energy gained would be used to pump protons across the cytoplasmic membrane. The ratchet could not only work on thermal fluctuations, but also on non-thermal fluctuations, including a constant current.

The flagella of many extant microorganisms could be part of a flagellar ratchet. Sessile organisms in environments with a local strong current are considered good prospects. Macroscopic algae such as *Laminaria* that grow in tidal currents (Chapman, 1962) may have emerged from large flagellar ratchets by the acquisition of chloroplasts. Use of flagellar ratchets is an alternative to heterotrophy as energy source in colourless algae (Pringsheim, 1963).

Until the Cambrian, when grazers appeared, the ocean floor was covered with filamentous organisms (Seilacher, 1999). Some of these filamentous organisms may, similar to the flagellar ratchet, have gained free energy from the current flow above the ocean floor. In the same way, sessile organisms on the ocean floor of the Jovian moon Europa could, as previously proposed (Schulze-Makuch and Irwin, 2002), by means of their flagella gain free energy from the uptake of warm vent water in large vacuoles or from the current in the surrounding water. With the flagellum sticking out in the warm vacuole or the moving ocean water, the flagellar ratchet could support both mechanisms.

Other environments of interest include the human body. Near heart valves the flow rate of the blood must be especially high. Because of their pathogenicity, the fastidious bacteria that cause endocarditis (Beynon et al., 2006) are prospects for the presence of a flagellar ratchet. A stimulation of their growth by the current flow should be experimentally observable; a dependence could explain their fastidiousness and localization on the valve.

*The reversed flagellar ratchet* When in Feynman's ratchet the pawls have a higher temperature than the vanes, the pawls set the vanes in motion (Feynman et al. 1963). A high temperature of the MotA/MotB proteins could similarly cause the flagellum to rotate. This 'reversed flagellar ratchet' must be distinguished from the flagellar pump or the flagellar motor. The organism would not change position— to profit from the thermal gradient it has to remain sessile. When a collection of reversed flagellar ratchets is present on the surface of a cell, a circulation in the water around the collection will be induced. The water circulation required for filter feeding (Jørgensen, 1975) may have emerged from such a collection above a biofilm of microorganisms on top of a submarine hydrothermal vent.

---

[5] Except for its much smaller size, the structure would resemble a wind turbine hanging in a balloon (Ferguson, 2005).

*Previously proposed thermal and Brownian ratchets*  Several studies have been published (Reimann and Hänggi, 2002) on 'thermal ratchets' and 'Brownian ratchets', notions also derived from Feynman's ratchet. Therein the ratchet is compared to a periodic potential, and the pawl to a particle moving in this potential (Feynman et al. 1963). There is some overlap with the ideas presented in this study, although the relevant biochemical context in these previous studies consists of filamentous proteins such as actin or tubulin, the linkage of monomers yielding the spatial periodicity. Some studies investigate the effect of thermal cycling (Bao, 1999; Li, 1997; Reimann et al., 1996; Zhang and Chen, 2008), similar as proposed here for the flagellar pump. These notions may be relevant for the emergence of contractile proteins and molecular motors.

## 4.6 The Flagellar Computer and the Turing Machine

A flagellated bacterium such as *E. coli* senses its environment and reacts to it by changing its direction of movement (Webre et al., 2003): this information collection and processing, followed by action, demonstrates its cognition capability. A bacterium is a computer. Because the flagellar motor contains all the proteins involved in chemotaxis (Berg, 2003), we consider whether the novel point of view given by thermosynthesis permits an explanation of the emergence and the mechanism of the 'flagellar computer'. ATP Synthase, the flagellar pump and the flagellar ratchet could generate the free energy that a computer in practice requires[6] (Bennet, 1982).

First we consider clocks and temporal periodicity—computation involves repetitive activity. Periodic phenomena must have played a key role during the origin of life (Muller, 1995). Moreover, any complex and composed system will require synchronization of its parts, which can be implemented by a clock. In today's personal computers, for example, the high-frequency alternating voltage, generated by a piezoelectric crystal, functions as the clock that synchronizes the parts. A previous study (Muller, 1996) proposed that early thermosynthesizing organisms similarly used thermal cycling to synchronize spatially separated cellular activities. Later in evolution, periodic $Ca^{2+}$-inflow would have taken over this role of thermal cycling in effecting internal cellular synchrony (Muller, 1996).

It seems most natural to associate gear-containing and dissipative clocks with complex, highly damped rotating devices such as ATP Synthase and the flagellar motor. One flagellar motor complex in a bundle of active flagellar motors could start to function as 'flagellar clock' or flagellar computer, when contact with the other active flagellar motors forces it to rotate and to drive the escapement of a clock.[7]

The ratchet in Fig. 15 has the same shape as the crown wheel of a *verge* escapement (Headrick, 1997) (Fig. 20A). Here the axis of the verge is parallel to the plane of the ratchet. In the *anchor* escapement the axes of the anchor and the rotor are parallel (Fig. 20B). In a flagellar motor a similar back and forth movement could occur and might be coupled to a biological clock. Several authors have proposed escapement mechanisms in biological motors (Huxley, 1981; Lowry and Frasch, 2005; Schilstra and Martin, 2006). Chemotaxis involves the delay or acceleration of tumbling. It is plausible that the emergence of the bacterial flagellar computer started by the similar delay or acceleration of a flagellar clock.

---

[6] Non-dissipative or reversible computing is theoretically feasible (Bennett, 1982), but no implementations have yet been proposed; a quantum computer (Benioff, 1982) could also be reversible.
[7] An escapement of a clock converts a rotary motion of the rotor (driven by a spring or weight) into a back-and-forth motion (Headrick, 1997).

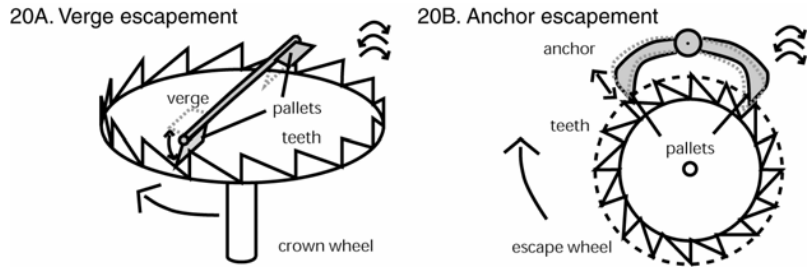

Fig. 20. *The verge and anchor escapements* Escapements are applied in clocks. In the verge escapement the teeth of a rotating crown wheel periodically lift a pallet connected to the verge; the other pallet then impedes the movement of the teeth at the other end of the verge. In this way the sought-for periodic back-and-forth motion is induced. In the anchor escapement the anchor performs a back-and-forth motion as well, the difference being that here the rotation axis of the anchor is parallel to the axis of the escape wheel. Similar mechanisms may be operating in rotating biomolecular devices such as ATP Synthase and the flagellar motor, permitting them to function as clocks as well.

A Turing machine is a theoretical, mathematical device first used to investigate the limits of algorithms, or more generally, of mathematics itself (Petzold, 2008; Turing, 1937). It is imagined to consist of (1) a *paper tape*— infinite in both directions—that can be read, written on and be moved forwards and backwards by (2) the *head*, which can assume a finite number of *states*, including an *initial state*, and (3) a *table*, that—using as input only the current *head state* and the symbols on the tape directly under the head—allows the head to write on the tape and move over it, in addition to changing its own state. Turing machines can be used to calculate many (but not all (Chaitin, 2007)) transcendental numbers such as $e$ and $\pi$.

Turing proved that it is possible to obtain a *description* on paper tape of *any* Turing machine that can calculate a certain transcendental number, so that a special Turing machine, the '*Universal Turing Machine'* (UTM), can calculate this transcendental number using this description. According to the Church-Turing thesis (Kleene, 1967), theoretical computers based on different principles still have the same calculation ability as the UTM, which, because of this universality, constitutes a suitable reference. Physical devices based on the UTM have had a spectacular success in the nowadays ubiquitous electronic computer.

Many physical implementation of Turing machines have been proposed, but many are also impractical because of low calculation speed. They may take many cycles to halt, which leads to the notion of a 'Turing tar-pit' (Perlis, 1982):

Beware of the Turing tar-pit in which everything is possible but nothing of interest is easy.

Halting is neither assured nor predictable (Chaitin, 2007; Turing, 1983). Physical devices with computation capabilities different from the Turing machine may also be useful (Stepney, 2008).

It follows from statistical mechanics that data storage of a bit cannot be 100% accurate (Bennett, 1979). More generally, errors are inherent to computer systems. According to Von Neumann (1956):

Error is viewed, therefore, not as an extraneous and misdirected or misdirecting accident, but as an essential part of the process under consideration—its importance in the synthesis of automata being fully comparable to that of the factor which is normally considered, the intended and correct logical structure.

Our present treatment of errors is unsatisfactory and ad hoc. It is the author's conviction, voiced over many years, that error should be treated by thermodynamic methods, and be the subject of thermodynamical theory, as information has been, by the work of L. Szilard and C.E. Shannon.

Whereas for the Turing machine it is assumed that errors during processing are absent—it is after all a logical, theoretical device—errors are significant in practical computers, and must especially have been significant in the first biological computers. Although Danchin (1998) claims that the cell is a Turing machine, several differences between biological computation systems and Turing machines have been recognized (Stepney, 2008):

> Turing computation, *designed* computation, is about halting, computability and universality; it is symbolic, discrete and closed (pre-defined); it is deterministic and sequential (in the sense that probabilistic or parallel variants provide no additional computation power); . . . its calculations are exceedingly fragile to small changes or errors. On the other hand, biological computation, *found* computation, is about not halting (halting equates system death); it is (mostly) nonsymbolic, continuous and open (constantly adapting and evolving due to the continual flow of matter, energy and information through the system); it is essentially stochastic and massively parallel; . . . and is robust to many classes of errors.

We consider the Turing machine at least partly applicable over the whole size range of organisms, from microorganisms such as bacteria to macroscopic organisms like ourselves that have a brain. Key questions on biological computers then concern their specifics: *How do they work, where are they located in the cell, how are programs physically realized, how did the machines evolve, what constitutes the difference between Turing and biological computation?* Previous investigators have analysed the link between the flagellar computer and Turing machines only in general terms (Berg, 1996), and the extensive literature on the link between the Turing machine and the human brain that starts with Turing (1938) himself gives little anatomical specifics.

Several correspondences between the flagellar motor and the Turing machine exist:
1. the ring of 26 MotA/MotB combinations in the flagellar motor can be linked to the paper tape; a similar number of switch complexes (FliG/FliM/FliN combinations) is present. The conformational state of a MotA/MotB combination or switch complex would stand for one bit. The motor is also able to read these conformational states while it turns: data can therefore be stored and read;
2. the switch complex can be linked to the Turing machine's head;
3. the head states to the current PMF, the conformations of the switch complex (CCW and CW), and the conformations of the flagellum —in addition to the CCW and CW state, additional changes in its shape are observed as well (Turner et al., 1996). The state of tumbling is linked to the Turing machine's initial state;
4. the mechanical interaction between the proteins of the flagellar motor would implement the table.

It is consequently postulated that the functions of information processing, storage and action during chemotaxis are implemented by a flagellar computer of which the structure, but not the function, is identical to the flagellar motor. During a run, a single flagellar ratchet in a rotating bundle of flagellar motors functions as flagellar computer, with the rest of the rotating bundle forcing the ratchet/computer's rotation. The ratchet writes down in its set of MotA/MotB proteins the state of the environment during the run, as indicated by the absence of the CheY-P protein. When the number of set MotA/MotB proteins in the ratchet reaches the threshold, the system (1) *resets these proteins* (cf. the 'initial state' of a Turing machine), (2) tumbles, and (3) starts another run.

Chemotaxis demonstrates that alternatives in computing to the Turing machine are possible and that nature has discovered them.[8] Proposed mechanisms for the flagellar computer could be confirmed by observation and experiment, in which conformational states of proteins are tracked by fluorescence or thermocoloration (Tomizaki and Mihara, 2007).

---

[8] Berg (1996) has drawn attention to the similarity of tumbling in bacteria and human thinking:
> Is a similar strategy fundamental to the operation of our brains? Does the sort of irritability that enables *E. coli* to change directions enable us to change our minds? Is there a secret here, learned early in evolution, that helps us think?

In addition to overcoming hardware and software errors, automated semi-periodic resetting overcomes the halting problem of the Turing machine. The name 'reset machine' is proposed for a computer with this reset capability. *A regular reset should be seen not as a rough method—such as effected by a Windows user on a personal computer—for correcting hardware or software errors, but as a characteristic of a computer architecture that differs from the Turing machine.* The reset machine accords with the recently formulated requirements for novel computing paradigms (Teuscher et al., 2008).

The model leaves open how the program is stored physically in the reset machine, and how it is reloaded. Moreover, a reset machine will function in a way different from a Turing machine, or even a regular computer: it is a distinct kind of automata. Upon a reset, data may have to be saved in stored memory. This may introduce errors in stored memory, and such damage may become significant in the long run, causing a kind of aging.

Webre *et al.* (2003) attribute to a bacterium that performs chemotaxis a 'bacterial nanobrain' situated at the site of the membrane receptors, instead of in the flagellar motor. Additional types of signal transduction, and more complex phenomena, on longer timescales, have also been interpreted as indicating prokaryotic intelligence (Ben-Jacob, 1998, 2006; Hellingwerf, 2005).

A related issue is the demonstrated presence in microorganisms of several types of memory different from regular DNA expression (Casadesús and D'Ari, 2002; Wolf et al., 2006). One could speak of computing when this memory is not simply the result of hysteresis of a non-linear reaction to environmental conditions, but can also be actively set by the organism. In a bacterium a network or maybe even a hierarchy of such, slightly coupled, data processing mechanisms may be present in addition to the bacterial flagellar computer.

## 4.7 Emergence of the Eukaryotic Cell

Until the nineties the large difference between the prokaryotic and the eukaryotic cell seemed conceptually insurmountable, but then many attributes considered typical of eukaryotes were also found in prokaryotes. Vellai and Vida (1999) list the nucleus, cytoskeleton, mitosis-supporting apparatus, internal membrane, linear chromosome, introns, eukaryotic-gene regulation, multiple genomes and large size. Remaining attributes characteristic of eukaryotes are identified as (1) an energy-producing cell organelle, (2) endocytosis capability, and (3) an ~ 100 000 times larger genome: whereas prokaryotes genomes typically contain 0,6 – 9,5 million bases, eukaryotic genomes start at 140 000 million bases. The large difference in the flagellar motors can be added (Margulis, 1970, 1981) to these characteristics. The bacterial flagellar motor works on the PMF, whereas the larger eukaryotic flagellar motor works on ATP. The rotating bacterial flagellum may be present on its own, or be part of a bundle. The 'bundle' of 9 + 2 microtubules inside the eukaryotic flagellum bends, a mechanism that may more easily scale up, as there seems to be a lower chance of entanglement of the longer microtubules.[9]

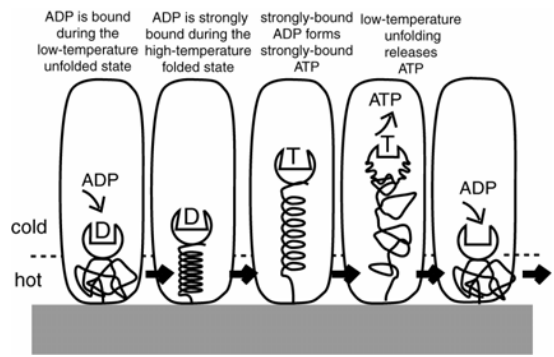

---

[9] The rotation of the central pair of microtubules observed in several eukaryotes (Mitchell and Nakatsugawa, 2004; Omoto and Kung, 1980) may be a relic of prokaryotic flagellar rotation.

Fig. 21. *Proposed emergence of muscle from a thermotether inside a cell* Fig. 2 depicts free energy gain from a thermotether *outside* a cell. A thermotether with an ATP generating FP attached to it operating *inside* a cell is also easily imagined. Reversal of such a system could lead to the emergence of muscle when an ATP-consuming FP evolved into myosin and the thermotether into actin.

The emergence of muscle can be explained similar to the emergence of the bacterial flagellar motor. A global glaciation again yielded the long evolutionary time required for the emergence of a thermosynthesis driven process that started with a thermotether moving back-and-forth in the thermal gradient *inside* a cell (Fig. 21). FPs attached to this internal thermotether generated condensation products such as ATP and GTP. Partial steps of the dissipative mechanisms of these proteins reversed at the end of the glaciation. The thermotether/FP combination would have turned into the actin/myosin couple of muscle. The source of the ATP required for muscle would have been respiration.

    The bacterial flagellar motor runs on protons pumped by the respiratory chain, which is situated in the cytoplasmic membrane, the only membrane in the prokaryotic cell. In the bacterium, respiration generates the free energy for both the flagellar motor and for metabolism. The demands of these two systems may conflict. Their separation seems advantageous and feasible by endosymbiosis of the progenitor of the ATP-producing mitochondrion (Martin and Müller, 1998). The ancient host of the mitochondrion, a subject of extensive debate in the biological literature (Andersson et al., 2003; Cavalier-Smith, 2006), is thus identified as a flagellated bacterium.

    By this endosymbiosis a division of labour was obtained that has remained present in organisms to today. The host transports substances to and from the mitochondria, which in turn generate the ATP with which the host runs this transport. A key aspect of the evolution of animals then concerns the evolution of transport. It would have started simple by a flagellar motor that ran on ATP, and would have ended with our complex systems for circulation, locomotion, and food and oxygen uptake.

    The mechanism of the prokaryotic flagellar ratchet proposed above is based on the rotating prokaryotic flagellum. The progenitor of the non- or less rotating eukaryotic flagellum may, in a thermal gradient, have generated ATP by a similar thermally induced bending: this progenitor would have constituted a eukaryotic flagellar ATP-generator that corresponded with the prokaryotic flagellar pump. Today's eukaryotic flagellum comprises several microtubule/dynein couples, and can be compared to a bundle of bacterial flagellar motors enclosed in a membrane (Fig. 3F). The growth of microtubules by addition of the tubulin monomer resembles the growth of the bacterial flagellum by the addition of the FliC monomer to the flagellum's tip.

    Emergence of an ATP-generating *eukaryotic flagellar ratchet* analogous to the proton-pumping bacterial flagellar ratchet seems feasible. Because of its bending it would resemble the invar-based circuit breaker used in car signalling lights mentioned in section 2. The notion should be kept in mind when investigating choanoflagellates and the similar choanocytes present in sponges. Protist ciliates connected to a substrate show quick contractions of the tether/stalk even in the absence of a thermal gradient, which hints at the presence of a ratchet (Moriyama et al., 1998).

    Another prospect for the eukaryotic flagellar ratchet is the flagellar cilium with its whip-like movement (Brennen and Winet, 1977). When present in a boundary layer it may span a thermal gradient. The flagellum's tip would have been thermally cycled during its movement, which therefore may have sustained a thermal relaxation oscillation. This mechanism may apply to the cilia in the bronchi of the lung where the temperature of inhaled air differs from the body temperature.

    Movement driven by the temperature difference across the base of the cilium can also be imagined: by analogy to the reversed prokaryotic flagellar ratchet one derives the notion of

the *reversed eukaryotic flagellar ratchet*,[10] in which the high temperature of the cilium would drive the movement of the eukaryotic flagellum.

*Eukaryotic complexity*  The task of resolving how the complex information processing and control occurring in the eukaryotic cell are physically implemented seems daunting. The unknown structure of an complex system may however be obtained by modelling its evolution: first establish what it did in the beginning and then consider the additions. Man-made complex systems such as airplanes—including their components such as jet engines— or computers (mainframe or personal) can be understood from their function. In contrast, it may be difficult to understand complex structures by the properties of their parts because of the redundancy in the numerous physicochemical properties that a part can have: the relevance of the properties of a part can only become clear from its function.

DNA (Adleman, 1994 ; Benenson et al., 2001), RNA (Faulhammer et al., 2000) and proteins (Bray, 1995; Unger and Moult, 2006) can be used for *in vitro* non-biological computation. We consider an evolutionary path from a *prokaryotic reset* to a *eukaryotic reset machine*. If in addition to *in vitro* computation, *in vivo* computation by biological macromolecules could be demonstrated, the hardware involved would plausibly descend from the bacterial flagellar computer. The eukaryotic computer may then, also plausibly, be found near the flagellum, i.e. underneath the basal body of the eukaryotic cilium and near the pair of centrioles in the middle of the centrosome (Kellogg et al., 1994). The cilium oscillates (Aoyama and Kamiya, 2005; Shingyoyi et al., 1998), and could therefore function as the clock of a computer. Recent studies indicate a central role in the control of the cell (Pan and Snell, 2007; Pazour and Witman, 2003; Quarmby and Parker, 2005), including tumour suppression (Mans et al., 2008).[11] Jekely and Arendt (2006) state:

> The evolution of the motile cilium was a major innovation during eukaryogenesis that, in synergy with endomembranes, phagocytosis, mitochondria and novel genome organisation, triggered the rapid diversification of the eukaryotes.

The proposed descent of the eukaryotic cilium from the bacterial flagellar motor contrasts with the notion that it was a eukaryotic innovation (Jekely and Arendt, 2006):

> Both autogenous and symbiotic scenarios have been put forward to explain the evolutionary origin of the cilium. Autogenous scenarios propose that the cilium evolved by the duplication and divergence of pre-existing components of the eukaryotic cell whereas in symbiotic scenarios, the merger of a motile Spirochete and a host cell is thought to have given rise to the organelle.

Several physical models for the function of the microtubules in the cilium have been proposed. Some invoke classical conformational changes in tubulins (Craddock and Tuszynski, 2007; Hameroff and Watt, 1982; Tuszynski et al., 1995), others invoke quantum processes including quantum computation (Craddock and Tuszynski, J.A., 2007; Hameroff, 1998). Some models comprise temperature-dependent conformational transitions (Tuszynski et al., 1995). The structures used in these models may be easily linked to the prokaryotic and eukaryotic flagellar computers proposed here.

The animal brain, a biological computer based on the action potential of the cell membrane of the neuron, differs substantially from the flagellar computer. A following study will give a model, based on the thermal diffusion potential (Agar, 1963), for the emergence of the nerve in a hydrothermal vent later in evolution at the onset of the Cambrian. Nerve-based computing would have been an add-on to an already functional eukaryotic flagellar computer that made use of the cilium.

---

[10] A progenitor of goose-flesh?

[11] In a transformed (tumour) cell the cellular computer could contain a changed program that results in a differently executed cell cycle: in that case cancer would be associated with a software bug. The difference between the normal and the transformed cell would then be very small, and hard to exploit by chemotherapy.

*Timing* This study identifies macroevolution with global glaciations during which life would have fallen back to thermosynthesis. Tentatively, the 2,9 Ga glaciation is associated with the emergence of the prokaryotic flagellar motor. Good dates for the first reported eukaryote fossils are 1,8 Ga (Knoll et al., 2006) and 2,1 Ga (Han and Runnegar, 1992) and their emergence is associated with the preceding global glaciation of 2,2 Ga. During the next global glaciations—the Snowball Earths of the late Proterozoic that start at 0,7 Ga—larger versions of the thermotether and novel thermosynthesizing organs based on different relaxation oscillations would have emerged that supported movement of larger volumes of water. These novel organs will be the subject of following studies.

## 5. Discussion

A key problem in biophysics is the question of how life overcame the overdamping of movement on the atomic scale (Purcell, 1977). The proposed simple solution, the thermotether, may have worked in the past in a thermal gradient, and may still be operating in thermal gradients in the world around us.

   Additional support is given to the notion that during its origin and early evolution, life did not move actively but was carried along by convection currents. Later the thermal gradients near hydrothermal vents permitted the emergence of directed motion based on simple relaxation oscillations that overcame the damping due to a small size. These processes gave thermosynthesizers a selective advantage during global glaciations and when these glacials ended, the reappearance of photosynthesis and respiration yielded the required energy for movement by mechanisms based on reversal of the previously acquired processes evolved from relaxation oscillations.

   In contrast to the standard Darwinian scenario for evolution, which assumes small incremental changes, the newly proposed scenario permits large changes: a mechanism for macroevolution (Gould, 1980, 1994; Stanley, 1979) driven by glaciations is obtained. Environmental fluctuations would have accelerated evolution (Boyle and Lenton, 2006; Kashtan et al., 2007). Martin (2005) has stated on the mechanism of the emergence of the eukaryotic nucleus that:

> A fundamental problem that is common to all ideas regarding the origin of the nucleus is that the underlying mechanism has to be plausible enough to have actually occurred, but at the same time so unlikely that it has only occurred once in four billion years . . . .This problem is severe. . . . It is the main reason that they are all coupled to a rare event in evolution, for example the origin of phagotrophy, a karyogenic symbiosis that occurred only in the eukaryotic lineage, or the origin of mitochondria. . . .

Such rare evolutionary events are therefore explained by macroevolution driven by long-lasting geological cycles of glacials and interglacials.

This study further increases the large explanatory power (Popper, 1963) of thermosynthesis. Seemingly disparate notions such as evolution, macroevolution, chemiosmosis, hydrothermal vents, global glaciations, heat engines, non-equilibrium thermodynamics, Feynman's ratchet-and-pawl mechanism, the bacterial flagellar motor and the eukaryotic cilium are combined in a unifying scenario that is verifiable by observation and experiment. The linkage of the flagellar motor to the Turing machine could have far-reaching consequences for the understanding of life: computation capability may be a key attribute of the so difficult to define living state (Danchin, 1998). Just as today's complex airplane can be understood as the result of the addition to a simple airplane of components that improve its function, the complexity of the eukaryotic cell may become understood as the result of the addition of specified components to a bacterial flagellar computer.

All theoretical modelling is easily called speculative, and this is in particular true for the modelling of the origin of life. The application of the notion of thermosynthesis to astrobiology (Muller, 2003) has been called 'science fiction' (Cockell and Westall, 2004). I do not agree. Given the important role of heat engines in physics and engineering, and the absence of a reason that would preclude a similar role in biology, it follows that biological heat engines are possible and even plausible. They may not have been observed because biologists have not looked for them—most are unfamiliar with their possibility, and the eye can in general only see what is already in the mind. The absence of progress in solving the problem of the origin of life means that a fundamental idea has probably been overlooked. This overlooked idea would not necessarily  have to be difficult and complex, but may instead be simple. The biological heat engine is a most suitable candidate for the overlooked fundamental idea that can explain the origin of life.

### 6. Conclusion

A simple theoretical cell organelle, the thermotether, is described that allows a microorganism to sustain a relaxation oscillation that generates free energy in the thermal gradient above submarine hydrothermal vents. This thermotether may have played a key role during early evolution. A modified thermotether that worked on a pH gradient instead of on a thermal gradient is a plausible progenitor of the proton translocation subunits of ATP Synthase and the prokaryotic flagellar motor. The flagellar pump resembles Feynman's ratchet-and-pawl mechanism. A model can be given for the emergence of a Turing-like machine that plausibly functions in prokaryotic and eukaryotic cells. The notion of the thermotether further increases the explanatory power of thermosynthesis in biology.

### Acknowledgements
Ronald Wouterson for a referral to the use of bimetal in car-signalling lights; Jaap Glas, Alice van der Laan and Amy E. Kelly for their suggestions for improving the manuscript.

### References

Adleman, L.M., 1994. Molecular computation of solution to combinatorial problems. Science 266, 1021-1024.

Agar, J.N., 1963. Thermogalvanic cells. Adv. Electrochem. Electroeng. 3, 31-121.

Andersen, O.S., Koeppe, R.E., 2007. Bilayer thickness and protein function: an energetic perspective. Annu. Rev. Biophys. Mol. Struct. 36, 107-130.

Andersson, S.G., Karlberg, O., Canback, B., Kurland, C.G., 2003. On the origin of mitochondria: a genomics perspective. Phil. Trans. R. Soc. B. 358, 165-177.

Angevine, C.M., Herold, K.A.G., Fillingame, R.H., 2003. Aqueous access pathways in subunit $a$ of rotary ATP synthase extend to both sides of the membrane. Proc. Natl Acad. Sci. USA 100, 13179-13183.

Aoyama, S., Kamiya, R., 2005. Cyclical interactions between two outer doublet microtubules in split flagellar axonemes. Biophys. J. 89, 3261-3268.

Armstrong, J.C., Wells, L.E., Gonzales, G., 2002. Rummaging through Earth's attic for remains of ancient life. Icarus 160, 183-196.

Azzone, G.F., Pozzan, T., Massari, S., Bragadin, M., 1978a. Proton electrochemical gradient and rate of controlled respiration in mitochondria. Biochim. Biophys. Acta 501, 296-306.

Azzone, G.F., Pozzan, T., Massari, S., 1978b. Proton electrochemical gradient and phosphate potential in mitochondria. Biochim, Biophys. Acta 501, 307-316.

Azzone, G.F., Pozzan, T., Massari, S., 1978c. Proton electrochemical gradient and phosphate potential in submitochondria particles. Biochim. Biophys. Acta 501, 317-329.

Bada, J.L., Bigham, C., Miller, S.L., 1994. Impact melting of frozen oceans on the early Earth: implications for the origin of life. Proc. Natl Acad. Sci. USA 91, 1248-1250.

Bada, J.L., Lazcano, A., 2002. Origin of Life, some like it hot—but not the first biomolecules. Science 296, 1982-1983.

Bahn, P.R., Pappelis, A., Grubbs, R., 2004. The role of heat in the origin of life. In: Seckbach, J., (Ed.), Life in the Universe. Kluwer, Dordrecht, The Netherlands, pp. 119-120.

Baker, J.E., 2004. Free energy transduction in a chemical motor model. J. Theor. Biol. 228, 467-471.

Bao, J.-D., 1999. Directed current of Brownian ratchet randomly circulating between two thermal sources. Physica A 273, 286-293.

Baross, J.A., Hoffman, S.E., 1985. Submarine hydrothermal vents and associated gradient environments as sites for the origin and evolution of life. Orig. Life 15, 327-345.

Bekker, A., Holland, H.D., Wang, P.-L., Rumble, D., Stein, H.J., Hannah, J.L., Coetzee, L.L., Beukes, N.J., 2004. Dating the rise of atmospheric oxygen. Nature 427, 117-120.

Benenson, Y., Paz-Elizur, T., Adar, R., Keinan, E., Livneh, Z., Shapiro, E., 2001. Programmable and autonomous computing machine made of biomolecules. Nature 414, 430-434.

Benioff, P.A., 1982. Quantum mechanical models of Turing Machines that dissipate no energy. Phys. Rev. Lett. 48, 1581-1585.

Bennett, C.H., 1973. Logical reversibility of computation. IBM J. Res. Dev. 6, 525-532.

Bennett, C.H., 1979. Dissipation-error tradeoff in proofreading. BioSystems 11, 85-91.

Bennett, C.H., 1982. The thermodynamics of computation—a review. Int. J. Theor. Physics 21, 905-940.

Ben-Jacob, E., 1998. Bacterial wisdom. Physica A 249, 553-557.

Ben Jacob, E., 2006. Seeking the foundations of cognition in bacteria: from Schrödinger's negative entropy to latent information. Physica A 359, 495-524.

Berg, H.C., 1996. Touring machines. Current Biology 6, 624.

Berg, H.C., 2003. The rotary motor of bacterial flagella. Annu. Rev. Biochem. 72, 19-54.

Beynon, R.P., Bahl, V.K., Prendergast, B.D., 2006. Infective endocarditis. Br. Med. J. 333, 334-339.

Bisschoff, J.L., Rosenbauer, R.J., 1984. The critical point of and two-phase boundary of seawater. 200-500°C. Earth Plan. Sc. Lett. 68, 172-180.

Bisschoff, J.L., Rosenbauer, R.J., 1988. Liquid-vapor relations in the critical region of the system NaCl-H$_2$O from 380 to 415°C: A refined determination of the critical point and the two-phase boundary of seawater. Geochim. Cosmochim. Acta 52, 2121-2126.

Black, S., 1970. Pre-cell evolution and the origin of enzymes. Nature 226, 754-755.

Boyer, P.D., 1979. The binding change mechanism of ATP synthesis. In: Lee, C.P., Schatz G., Enster, L., (Eds), Membrane Bioenergetics. Addison-Wesley, London, pp. 461-479.

Boyer, P.D., 1997. The ATP synthase — a splendid molecular machine. Annu. Rev. Biochem. 66, 717-749.

Boyle, R.A., Lenton, T.M., 2006. Fluctuation in the physical environment as a mechanism for reinforcing evolutionary transitions. J. Theor. Biol. 242, 832-843.

Brandts, J.F., 1967. Heat effects on proteins and enzymes. In: Rose A.H., (Ed.), Thermobiology. Academic Press, London, pp. 25-72.

Bray, D., 1995. Protein molecules as computational elements in living cells. Nature 376, 307-312.

Brennen, C., Winet, H., 1977. Fluid mechanics of propulsion by cilia and flagella. Annu. Rev. Fluid Mech. 9, 339-398.

Broda, E., 1975. The evolution of the bioenergetic processes. Pergamon, Oxford.

Brown, P.N., Terrazas, M., Paul, K., Blair, D.F., 2007. Mutational analysis of the flagellar protein FliG: sites of interaction with FliM and implications for organization of the switch complex. J. Bacteriol. 189, 305-312.

Buford Price, P., 2007. Microbial life in glacial ice and implications for a cold origin of life. FEMS Microbiol. Ecol. 59, 213-231.

Carnot, S., 1824. Réflexions sur la puissance motrice du feu et sur les machines propres a développer cette puissance. Bachelier, Paris.

Casadesús, J., D'Ari, R., 2002. Memory in bacteria and phage. BioEssays 24, 514-518.

Cavalier-Smith, T., 2006. Origin of mitochondria by intracellular enslavement of a photosynthetic purple bacterium. Proc. Roy. Soc. B 273, 1943-1952.

Chaitin, G.J., 2007. Thinking about Gödel and Turing. World Scientific, New Jersey.

Chapman, V.J., 1962. The Algae. MacMillan, London.

Chen, J.-W., Oliveri, P., Gao, F., Dornbos, S.Q., Li, C.-W., Bottjer, D.J. Davidson, E.H., 2002. Precambrian animal life: probable developmental and adult cnidarian forms from Southwest China. Devel. Biol. 248, 182-196.

Chng, C.-P., Kitao, A., 2008. Thermal unfolding simulations of bacterial flagellin: insight into its refolding before assembly. Biophys. J. 94, 3858-3871.

Chyba, C.F., 1993. The violent environment of the origin of life: progress and uncertainties. Geochim. Cosmochim. Acta 57, 3351-3358.

Cockell, C.S., Westall, F., 2004. A postulate to assess 'habitability'. Int. J. Astrobiol. 3, 157-163.

Cohen, B.A., Swindle, T.D., King, D.A., 2000. Support for the lunar cataclysm hypothesis from lunar meteorite impact melt ages. Science 290, 1754-1756.

Cohen-Ben-Lulu, G.N., Francis, N.R., Shimoni, E., Noy, D., Davidov, Y., Prasad, K., Sagi, Y., Cecchini, G., Johnston, R.M., Eisenbach, M., 2008. The bacterial flagellar switch complex is getting more complex, EMBO J. 27, 1134-1144.

Conway Morris, S., 2006. Darwin's dilemma: the realities of the Cambrian explosion. Proc. Trans. R. Soc. B. 361, 1069-1083.

Copley, S.D., 2003. Enzymes with extra talents; moonlighting functions and catalytic promiscuity. Curr. Opinion Chem. Biology 7, 265-272.

Corliss, S.B., Baross, J.A., Hoffman, S.E., 1981. An hypothesis concerning the relationship between submarine hot springs and the origin of life on Earth. Oceanologica Acta 4 (suppl) 59-69.

Corliss, S.B., Dymond, J., Gordon, L.I., Edmond, J.M., von Herzen, R.P., Ballard, R.D., Green, K., Williams, D., Bainbridge, A., Crane, K., van Andel, T.H., 1979. Submarine thermal springs on the Galápagos Rift. Science 203, 1073-1083.

Cornelis, G.R., 2006. The type III secretion injectisome. Nature Rev. Microbiol. 4, 811-825.

Craddock, T.J.A., Tuszynski, J.A., 2007. On the role of the microtubule in cognitive brain functions. NeuroQuantology 5, 32-57.

Cramer, W.A., Knaff, D.B., 1991. Energy Transduction in Biological Membranes. A Textbook of Bioenergetics. Springer-Verlag, New York.

Cuezva, J.M., Sánchez-Aragó, M., Sala, S., Blanco-Rivero, A., Ortega, A.D., 2007. A message emerging from development: the repression of mitochondrial β-F1-ATPase expression in cancer. J. Bioenerg. Biom. 39, 259-265.

Danchin, A., 1998. The Delphic boat: what genomes tell us. Harvard University Press.

Darwin, C., 1859. On the origin of species by means of natural selection. J. Murray, London.

DeMeis, L., 1989. Role of water in energy of hydrolysis of phosphate compounds—energy transduction in biological membranes. Biochim. Biophys. Acta 973, 333-349.

De Groot, S.R., Mazur, P., 1962. Non-equilibrium Thermodynamics. North-Holland Publishing, Amsterdam.

Dimroth, P., Von Ballmoos, C., Meier, T., 2006. Catalytic and mechanical cycles in F1-ATP synthases. EMBO Reports 7, 276-282.

Doebeli, M., Dieckmann, U., 2003. Speciation along environmental gradients. Nature 421, 259-264.

Engelmann, T.W., 1893. Über den Ursprung der Muskelkraft. Leipzig.

Evans, D.A., Beukes, N.J., Kirschvink, J.L., 1997. Low-latitude glaciation in the Palaeoproterozoic era. Nature 386, 262-266.

Faulhammer, D., Cukras, A.R., Lipton, R.J., Landweber, L.F., 2000. Molecular computation: RNA solutions to chess problems. Proc. Natl Acad. Sci. USA 97, 1385-1389.

Feniouk, B.A., Mulkidjanian, A.Y., Junge, W., 2005. Proton slip in the ATP synthase of *Rhodobacter capsulatus*: induction, proton conduction, and nucleotide dependence.

Biochim. Biophys. Acta 1706, 184-194.

Ferguson, F.D., 2005. Systems and methods for tethered wind turbins, U.S. Patent 7335000.

Feynman, R., Leighton, R., Sands, M., 1963. The Feynman Lectures on Physics, Reading, MA. Addison-Wesley.

Fick, A., 1893a. Einige Bemerkungen zu Engelmann's Abhandlung über den Ursprung der Muskelkraft. Arch. Ges. Physiol. 53, 606-615.

Fick, A., 1893b. Noch einige Bemerkungen zu Engelmann's Schrift über den Ursprung der Muskelkraft. Arch. Ges. Physiol. 54, 313-318.

Filliger, R., Reimann, P., 2007. Brownian gyrator, a minimal heat engine on the nanoscale. Phys. Rev. Lett. 99, 230602.

Florkin, M., 1972. A history of biochemistry. In: Florkin, M., Stotz, E.H., (Eds.), Comprehensive biochemistry. Vol. 30. Elsevier, Amsterdam, p. 244.

Flory, P.J., 1953. Principles of Polymer Chemistry. Cornell University.

Flory, P.J., 1956. Theory of elastic mechanisms in fibrous proteins. J. Am. Chem. Soc. 78, 5222-5235.

Fox, S.W., Dose, K., 1977. Molecular evolution and the origin of life. Marcel Dekker, New York.

Frauenfelder, H., Fenimore, P.W., Chen, G., McMahon, B.H., 2006. Protein folding is slaved to solvent motions. Proc. Natl Acad. Sci. USA 103, 15469-15472.

Friedel, M., Baumketner, A., Shea, J.-E., 2006. Effects of surface tethering on protein folding mechanisms. Proc. Natl Acad. Sci. USA 103, 8396-8401.

Gaeta, F.S., Bencivenga, U., Canciglia, P., Rossi, S., Mita, D.G., 1987. Temperature gradients and prebiological evolution. Cell Biochem. Biophys. 10, 103-125.

Gafni, A., Boyer, P.D., 1985. Modulation of stochiometry of the sarcoplasmic reticulum calcium pump may enhance thermodynamic efficiency. Proc. Natl Acad. Sci. USA 82, 98-101.

Galan, J.E., Wolf-Watz, H., 2006. Protein delivery into eukaryotic cells by type III secretion machineries. Nature 444, 567-573.

Giersch, C., Heber, U., Kobayashi, Y., Inoue, Y., Shibata, K., Heldt, H.W., 1980. Energy charge, phosphorylation potential and proton motive force in chloroplasts. Biochim. Biophys. Acta 590, 59-73.

Glaessner, M.F., 1984. The dawn of animal life: a biohistorical study. Cambridge University Press, Cambridge.

Gonzalo, J.A., 1976. Ferroelectric materials as energy converters. Ferroelectrics 11, 423-430.

Gough, D.O., 1981. Solar interior structure and luminosity variations. Sol. Phys. 74, 21-34.

Gould, S.J., 1980. Is a new and general theory of evolution emerging? Paleobiology 6, 116-130.

Gould, S.J., 1994. Tempo and mode in the macroevolutionary reconstruction of Darwinism. Proc. Natl Acad. Sci. USA 91, 9413-9417.

Gould, S.J., Eldredge, N., 1993. Punctuated equilibrium comes of age. Nature 366, 223-227.

Guillaume, C.E., 1897. Recherche sur les aciers au nickel. Dilatations aux temperature elevees; resistance electrique. Comptes Rendus Acad. Sci. 125, 235-238

Haase, R., 1963. Thermodynamik der irreversiblen Prozesse. Dietrich Steinkopf, Darmstadt.

Halliday, A.N., 2000. Terrestrial accretion rates and the formation of the Moon. Earth Plan. Science Lett. 176, 17-30.

Hameroff, S., 1998. Quantum computation in brain microtubules? The Penrose-Hameroff 'Orch OR' model of consciousness. Phil. Trans. Roy. Soc. Lond. A 356, 1869-1896.

Hameroff, S.R., Watt, R.C., 1982. Information processing in microtubules. J. Theor. Biol. 98, 549-561.

Han, T.M., Runnegar, B., 1992. Megascopic algae from the 2.1 billion-year-old Negaunee Iron Formation, Michigan. Science 257, 232-235.

Hardt, S.L., 1980. A measurable consequence of the Mitchell theory of phosphorylation. J. Theor. Biol. 87, 1-7.

Harold, F.M., 2001. Gleanings of a chemiosmotic eye. BioEssays 23, 848-855.

Hasegawa, E., Kamiya, R., Asakura, S., 1982. Thermal transitions in helical forms of *Salmonella* flagella. J. Mol. Biol. 160, 609-621.

Headrick, M.V., 1997. Clock and watch escapement mechanics.

Hellingwerf, K.J., 2005. Bacterial observations: a rudimentary form of intelligence. Trends Microbiol. 13, 152-158.

Hightower, K.E., McCarty, R.E., 1996. Influence of nucleotides on the cold stability of chloroplast coupling factor 1. Biochemistry 35, 10051-10057.

Hofmann, B.A., Farmer, J.D., Von Blanckenburg, F., Fallick, A.E., 2008. Subsurface filamentous fabrics: an evaluation of origins based on morphological and geochemical criteria, with implications for exopaleontology. Astrobiology 8, 87-117.

Hoffman, P.J., Kaufman, A.J., Halverson, G.P., Schrag, D.P., 1998. A neoproterozoic snowball Earth. Science 281, 1342-1346.

Hoh, S.R., 1963. Energy converter. U.S. Patent 3073974.

Holland, I.B., 2004. Translocation of bacterial proteins—an overview. Biochim. Biophys. Acta 1694, 5-16.

Hollerbach, R., 1996. On the theory of the geodynamo. Phys. Earth. Planet. Interiors 98, 163-185.

Horita, J., 2005. Some perspectives on isotope biosignatures for early life. Chem. Geology 218, 171-186.

Horneck, G., Stöffler, D., Ott, S., Hornemann, U., Cockell, C.S., Moeller, R., Meyer, C., De Vera, J.-P., Fritz, J., Schade, S., Artemieva, A., 2008. Microbial inhabitants survive hypervelocity impacts on Mars-like host planets: first phase of lithopanspermia experimentally tested. Astrobiology 8, 17-43.

Howard, J., 2001. Mechanics of motor proteins and the cytoskeleton. Sinauer, Sunderland, MA.

Hueck, C.J., 1998. Type III protein secretion systems in bacterial pathogens of animals and plants. Microb. Mol. Biol. Rev. 62, 379-433.

Hult, K., Berglund, P., 2007. Enzyme promiscuity: mechanism and applications. Trends Biotechn. 25 (5), 231-238.

Huston, D.L., Logan, G.A., 2004. Barite, BIFs and bugs, evidence for the evolution of the Earth's early atmosphere. Earth Planet. Sci. Lett. 220, 41-55.

Huxley, A.F., 1981. Reflexions on muscle. Princeton University Press.

Imada, K., Minamino, T., Tahara, A., Namba, K., 2007. Structural similarity between the flagellar type III ATPase FliI and $F_1$-ATPase subunits. Proc. Natl Acad. Sci. USA 104, 485-490.

Imai, E., Honda, H., Brack, A., Matsuno, K., 2005. Elongation of oligopeptides in a simulated submarine hydrothermal system. Science 283, 831-833.

Jagendorf, A.T., Uribe, E., 1966. ATP formation caused by acid-base transition of spinach chloroplasts. Proc. Natl Acad. Sci. USA 55, 170-177.

Jannasch, H.W., Wirsen, C.O., 1981. Morphological survey of microbial mats near deep-sea thermal vents. Appl. Environm. Microbiol. 41, 528-538.

Jao, S.C., Huang, L.F., Tao, Y.S., Li, W.S., 2004. Hydrolysis of organophosphate triesters by *Escherichia coli* aminopeptidase P. J. Mol. Catalysis B: Enzymatic 27, 7-12.

Jekely, G., Arendt, D., 2006. Evolution of intraflagellar transport from coated vesicles and autogenous origin of the eukaryotic cilium. BioEssays 28, 191-198.

Jørgensen, C.B., 1975. Comparative physiology of suspension feeding. Annu. Rev. Physiol. 37, 57-79.

Kalanetra, K.M., Huston, S.L., Nelson, D.C., 2004. Novel, attached, sulphur-oxidizing bacteria at shallow hydrothermal vents possess vacuoles not involved in respiratory nitrate accumulation, Appl. Environm. Microbiol. 70, 7487-7496.

Kashtan, N., Noor, E., Alon, U., 2007. Varying environments can speed up evolution. Proc. Natl Acad. Sci. USA 104, 13711-13716.

Kasting, J.F., Eggler, D.H., Raeburn, S.P., 1993. Mantle redox evolution and the oxidation state of the Archean atmosphere. J. Geol. 101, 245-257.

Kasting, J.F., Howard, M.T., 2006. Atmospheric composition and climate on the early Earth, Phil. Trans. R. Soc. B 361, 1733-1742.

Kasting, J.F., Ono, S., 2006. Palaeoclimates: the first two billion years. Phil. Trans. R. Soc. B 361, 917-929.

Kazlauskas, R.J., 2005. Enhancing catalytic promiscuity for biocatalysis. Curr. Opinion Chem. Biology 9, 195-201

Kelley, D.S., Karson, J.A., Blackman, D.K., Früh-green, G.L., Butterfield, D.A., Lilley, M.D., Olson, E.J., Schrenk, M.O., Roe, K.K., Lebon, G.T., Rivizzigno, P, the AT3-60 Ship Board Party, 2001. An off-axis hydrothermal vent field near the Mid-Atlantic Ridge at 30°N. Nature 412, 145-149.

Kellogg, D.R., Moritz, M., Albertz, B.M., 1994. The centrosome and cellular organization. Annu. Rev. Biochem. 63, 639-674.

Kettner, C., Bertl, A., Obermyer, G. , Slayman, C., Bihler, H., 2003. Electrophysiological analysis of the yeast V-type proton pump, variable coupling ratio and proton shunt. Biophys. J. 85, 3730-3738.

Khalil, M.A.K., 1999. Non-$CO_2$ greenhouse gases in the atmosphere. Annu. Rev. Energy Environm. 24, 645-661.

Kharecha, P., Kasting, J.F., Siefert J.L., 2005. A coupled atmosphere-ecosystem model of the Early Archean Earth. Geobiology 3, 53-76.

Khersonsky , O., Roodveldt, C., Tawfik, D.S., 2006. Enzyme promiscuity: evolutionary and mechanistic aspects. Curr. Opinion Chem. Biology 10, 498-508.

Kirschvink, J.L., 1992. Late Proterozoic, low-latitude global glaciation, The Snowball Earth. In: Schopf, J.W., Klein, C., (Eds.), The Proterozoic biosphere: A multidisciplinary study. Cambridge, Cambridge University Press, pp. 51-52.

Kleene, S.C., 1967. Mathematical logic. Wiley, New York.

Knoll, A.H., Javaux, E.J., Hewitt, D., Cohen, P., 2006. Eukaryotic organisms in Proterozoic oceans. Phil. Trans. R. Soc. B 361, 1023-1038.

Kojima, S., Blair, D.F., 2001. Conformational change in the stator of the bacterial flagellar motor, Biochemistry 40, 13041-13050.

Komissarov, G.G., 1990. Photosynthesis as a thermal process. In: Baltscheffsky, M., (Ed.), Current Research in Photosynthesis. Vol. 4. Kluwer, Deventer, pp. 107-110.

Kompanichenko, V., Frisman, E., Fishman, B., Savenkova, E., Shlufman, K., 2008. Exploration of thermodynamic fluctuations in the probable hydrothermal medium for the origin of life (on the example of Mutnovsky hydrothermal system in Kamchatka). Int. J. Astrobiology 6, 74.

Konings, W.N., Booth, I.R., 1981. Do the stoichiometries of ion-linked transport systems vary? Trends Biochem. Sci 6, 257-262.

Kramer, H.A., 1940. Brownian motion in a field of force and the diffusion model of chemical reactions. Physica 7, 284-304.

Larsen, S.H., Adler, J., Gargus, J.J., Hogg, R.W., 1974. Chemomechanical coupling without ATP: the source of energy for motility and chemotaxis in bacteria. Proc. Natl Acad. Sci. USA 71, 1239-1243.

Levin, L.A., 2002. Deep-ocean life where oxygen is scarce. Am. Scientist 90, 436-444.

Li, Y.-X., 1997. Transport generated by fluctuating temperature. Physica A 238, 245-251.

Liu, R., Ochman, H., 2007. Stepwise formation of the bacterial flagellar system. Proc. Natl Acad. Sci. USA 104, 7116-7121.

Lowry, D.S., Frasch, W.D. (2005) Interactions between βD372 and γ subunit N-terminal residues γ K9 and γS12 are important to catalytic activity catalyzed by *Escherichia coli* $F_1F_o$-ATP Synthase. Biochemistry 44, 7275-7281.

Machura, L., 2006. Performance of Brownian Motors. Thesis, University of Augsburg.

MacNab, R.M., 1979. An entropy-driven engine—the bacterial flagellar motor. In: Oplatka, A., Balaban, M., (Eds), Biological Structures and Coupled Flows. Academic Press, New York, pp. 147-160.

MacNab, R.M., 2003. How bacteria assemble flagella. Annu. Rev. Microbiol. 57, 77-100.

MacNab, R.M., 2004. Type III flagellar protein export and flagellar assembly. Biochim. Biophys. Acta 1694, 207-217.

Mans, D.A., Voest, E.E., Giles, R.H., 2008. All along the watchtower: Is the cilium a tumor suppressor organelle? Biochim. Biophys. Acta. In press. doi:10.1016/j.bbcan.2008.02.002.

Margulis, L., 1970. Origin of eukaryotic cells. New Haven, Yale University Press.

Margulis, L., 1981. Symbiosis in cell evolution. San Francisco Freeman.

Martin, W., 2005. Archaebacteria (Archaea) and the origin of the eukaryotic nucleus. Curr. Opin. Microbiol. 8, 630-637.

Martin, W., Müller, M., 1998. The hydrogen hypothesis for the first eukaryote. Nature 392, 37-44.

Martin, W., Russell, M.J., 2002. On the origins of cells: a hypothesis for the evolutionary transitions from abiotic geochemistry to chemoautotrophic prokaryotes, and from prokaryotes to nucleated cells. Phil. Trans. Roy. Soc. Lond. B 385, 59-85.

Mashimo, T., Hashimoto, M., Yamaguchi, S., Aizawa, S.-I., 2007. Temperature-hypersensitive sites of the flagellar switch component FliG in *Salmonella enterica* Serovar Typhimurium. J. Bacteriol. 189, 5153-5160.

Matsuno, K., 1995. Physics underlying the formation of protocells. J. Biol. Physics 20, 117-121.

Matsuno, K., 1996. A possible prebiotic heat engine, Origins Life Evol. Biospheres 26, 458.

Mawson, D., 1949. The late Precambrian ice-age and glacial record of the Bibliando dome. J. Proc. Roy. Soc. New South Wales, 82, 150-174.

McCall, G.J.H., 2006. The Vendian (Ediacaran) in the geological record: enigmas in geology's prelude to the Cambrian explosion. Earth-Science Reviews 77, 1-229.

Metzner, P., 1920. Die Bewegung und Reizbeantwortung der bipolar begeißelten Spirillen, Jahrbücher f. Wissensch. Botanik 59, 325-412.

Miller, S.L., Bada, J.L., 1988. Submarine hot springs and the origin of life. Nature 334, 609-611.

Miller, S.L., Lazcano, A., 1995. The origin of life—did it occur at high temperatures? J. Mol. Evol. 41, 689-692.

Mitchell, D.R., Nakatsugawa, M., 2004. Bend propagation drives central pair rotation in *Chlamydomonas reinhardtii* flagella. J. Cell Biol. 166, 709-715.

Mitchell, P., 1961. Coupling of phosphorylation to electron and proton transfer by a chemi-osmotic type of physiology. Nature 191, 144-148.

Mitchell, P., 1979. Keilin's respiratory chain concept and its chemiosmotic consequences. Science 206, 1148-1159.

Moriyama, Y., Hiyama, A., Asai, H., 1998. High-speed video cinematographic demonstration of stalk and zooid contraction of *Vorticella convallaria*. Biophys. J. 74, 487-491.

Mullen, J.G., Look, G.W., Konkel, J., 1975. Thermodynamics of a simple rubber-band heat engine. Am. J. Physics 43(4), 349-353.

Muller, A.W.J., 1983. Thermoelectric energy conversion could be an energy source of living organisms. Phys. Lett. 96 A, 319-321.

Muller, A.W.J., 1985. Thermosynthesis by biomembranes: energy gain from cyclic temperature changes. J. Theor. Biol. 115, 429-453.

Muller, A.W.J., 1993. A mechanism for thermosynthesis based on a thermotropic phase transition in an asymmetric biomembrane. Physiol. Chem. Phys. Medical NMR 25, 95-111.

Muller, A.W.J., 1995. Were the first organisms heat engines? A new model for biogenesis and the early evolution of biological energy conversion. Prog. Biophys. Mol. Biol. 63, 193-231.

Muller, A.W.J., 1996. The thermosynthesis model for the origin of life and the emergence of regulation by $Ca^{2+}$. Essays Biochem. 31, 103-119.

Muller, A.W.J., 2003. Finding extraterrestrial organisms living on thermosynthesis. Astrobiology 3, 555-564.

Muller, A.W.J., 2005a. Thermosynthesis as energy source for the RNA world: A model for the bioenergetics of the origin of life. BioSystems 82, 93-102.

Muller, A.W.J., 2005b. Photosystem 0, a proposed ancestral photosystem without reducing power that uses metastable light-induced dipoles for ATP synthesis. Published in the physics archives: www. arXiv.net/physics/0501050.

Muller, A.W.J., 2008. Emergence of animals during Snowball Earths from biological heat engines in the thermal gradient above submarine hydrothermal vents. Book of abstracts. XII ISSOL meeting. XV International conference on the origin of life. Florence, p. 184.

Muller, A.W.J., Schulze-Makuch, D., 2006a. Sorption heat engines: simple inanimate negative entropy generators. Physica A 362, 369-381.

Muller, A.W.J., Schulze-Makuch, D., 2006b. Thermal energy and the origin of life. Orig. Life Evol. Biosph. 36, 177-189.

Narbonne, G.M., 1998. The Ediacara biota: a terminal Neoproterozoic experiment in the evolution of life. GSA Today 8 (20), 1-6.

Narbonne, G.M., 2004. Modular construction of early Ediacaran complex life forms. Science 305, 1141-1144.

Neidhart, F.C., 1987. Chemical composition of *Escherichia coli*. In: Neidhart, F.C., Ingraham, J.L., Low, K.B., Magasanik, B., Schaechter, M., Umbarger, H.E. (Eds) *Escherichia coli* and *Salmonella typhimurium*: cellular and molecular biology. American Society for Microbiology, Washington DC, pp. 3-6.

Nelson, D.C., Wirson, C.O., Jannasch, H.W., 1989. Characterization of large, autotrophic *Beggiatoa* spp. abundant at hydrothermal vents of the Guaymas Basin. Appl. Environm. Microbiol. 55, 2909-2917.

Omoto, C.K., Kung, C., 1980. Rotation and twist of the central-pair microtubules in the cilia of *Paramecium*. J. Cell Biol. 87, 33-46.

Pan, J., Snell, W., 2007. The primary cilium: keeper of the key to cell division, Cell 129, 1255-1257.

Paster, E., Ryu, W.S., 2008. The thermal impulse response of *Escherichia coli*. Proc. Natl Acad. Sci. USA 105, 5373-5377.

Pazour, G.J., Witman, G.B., 2003. The vertebrate primary cilium is a sensory organelle. Curr. Op. Cell Biol. 15, 105-110.

Perlis, A.J., 1982. Epigrams on programming. http://luggage.bcs.uwa.edu.au/~michael/ /Perlis_Epigrams.html

Petzold, C., 2008. The annotated Turing: a guided tour through Alan Turing's historic paper on computability and the Turing machine. Wiley, Indianapolis.

Pirajno, F., van Kranendonk, M.J., 2005. Review of hydrothermal processes and systems on Earth and implications for Martian analogues. Aust. J. Earth Sciences 52, 329-351.

Popper, C.R., 1963. Conjectures and refutations: the growth of scientific knowledge. Routledge, London.

Pringsheim, E.G., 1963. Farblose Algen. Ein Beitrag zur Evolutionsforschung. Gustav Fischer, Stuttgart.

Pulvari, C.F., Garcia, F.J., 1978. Conversion of solar to electrical energy utilizing the thermodielectric effect. Ferroelectrics 22, 769-771.

Purcell, E.M., 1977. Life at low Reynolds number. Am. J. Physics 45, 3-11.

Quarmby, L.M., Parker, J.D.K., 2005. Cilia and the cell cycle? J. Cell Biol. 169, 707-710.

Rasmussen, B., 2000. Filamentous microfossils in a 3,235-million-year-old volcanogenic massive sulphide deposit. Nature 405, 676-679.

Reed, C., 2006. Boiling points. Nature 439, 905-907.

Reed, R.C., 2006. The Superalloys. Fundamentals and applications. Cambridge University Press.

Reimann, P., Bartussek, R., Häußler, R., Hänggi, P., 1996. Brownian motors driven by temperature oscillations. Phys. Lett. A 215, 26-31.

Reimann, P., Hänggi, P., 2002. Introduction to the physics of Brownian motors. Appl. Physics A 75, 169-178.

Reysenbach, A.-L., Cady, S.L., 2001. Microbiology of ancient and modern hydrothermal systems. Trends Microbiol. 9(2), 79-86.

Richardson, S.M., 1989. Fluid mechanics. Hemisphere Publishing, New York, p. 202.

Ritort, F., 2003. Work fluctuations, transient fluctuations of the second law and free-energy recovery methods. Poincaré Seminar 2, 195-229.

Roberts, P.H., Soward, A.M., 1992. Dynamo theory. Annu. Rev. Fluid Mech. 24, 459-512.

Rosing, J., Slater, E.C., 1972. The value of $\Delta G^0$ for the hydrolysis of ATP. Biochim. Biophys. Acta 267 275-290.

Russell, M.J., 2007. The alkaline solution to the emergence of life: energy, entropy and early evolution. Acta Biotheor. 55, 133-179.

Russell, M.J., Arndt, N.T., 2005. Geodynamic and metabolic cycles in the Hadean. Biogeosciences 2, 97-111.

Russell, M.J., Hall, A.J., 1997. The emergence of life from iron monosulphide bubbles at a submarine hydrothermal redox and pH front. J. Geol. Soc. Lond. 154 , 377-402.

Ryder, G., 1990. Lunar samples, lunar accretion and the end of the early bombardment of the Moon. Eos 71, 322-323.

Sagan, C., Mullen, G., 1972. Earth and Mars: evolution of atmospheres and surface temperatures. Science 177, 52-56

Santelli, C.M., Orcutt, B.N., Banning, E., Bach, W., Moyer, C.L., Sogin, M.L., Staudigel, H., Edwards, K.J., 2008. Abundance and diversity of microbial life in ocean crust. Nature 453, 653-657.

Savarino, G., Fisch, M.R., 1991. A general physics laboratory investigation of the thermodynamics of a rubber band. Am. J. Physics 59(2), 141-145.

Schilstra, M.J., Martin, S.R., 2006. An elastically tethered viscous load imposes a regular gait on the motion of myosin-V. Simulation of the effect of transient force relaxation on a stochastic process. J. Royal Soc. Interface 3, 153-165.

Schlodder, E., Gräber, P., Witt, H.T., 1982. Mechanism of phosphorylation in chloroplasts. Top. Photosynth. 4, 105-175.

Schmitt, R., 2003. Helix rotation of the flagellar rotary motor. Biophys. J. 85, 843-849.

Schopf, J.W., 2006. Fossil evidence of Archaean life. Phil. Trans. R. Soc. B 361, 869-885.

Schubert, G., Turcotte, D.L., Olson, P., 2001. Mantle Convection in the Earth and Planets. Cambridge University Press, Cambridge.

Schulz, H.N., Jørgensen, B.B., 2001. Big bacteria. Annu. Rev. Microbiol. 55, 105-107.

Schulze-Makuch, D., Irwin, L.N., 2002. Energy cycling and hypothetical organisms in Europa's ocean. Astrobiology 2, 105-121.

Schwartzman, D., Caldeira, K., Pavlov, A., 2008. Cyanobacterial emergence at 2.8 Gya and greenhouse feedbacks. Astrobiology 8, 187-203.

Seilacher, A., 1999. Biomat-related life styles in the Precambrian. Palaios 14, 86-93.

Shingyoyi, C., Higuchi, H., Yoshimura, M., Katayama, E., Yanagida, T., 1998. Dynein arms are oscillatory force generators. Nature 393, 711-714.

Shock, E.L., 1990. Geochemical constraints of the origin of organic compounds in hydrothermal systems. Orig. Life. Evol. Biosph. 20, 331-367.

Silverman, M., Simon, M., 1974. Flagellar rotation and the mechanism of bacterial motility. Nature 249, 73-74.

Simpson, E.D., Finney, W.H., 1930. The effect of acids and their salts on heat contraction of rat tendon. Q. J. Exp. Physiol. 20, 115-123.

Singh, R., 1989. A rubber heat engine for ground water irrigation in India. Agriculture, Ecosystems and Environment 25, 271-278.

Sklar, A.A., 2005. A numerical investigation of a thermodielectric power generation system. Thesis, Georgia Institute of Technology.

Slack, J.F., Grenne, T., Bekker, A., Rouxel, O.J., Lindberg, P.A., 2007. Suboxic deep seawater in the late Paleoproterozoic: evidence from hematitic chert and iron formation related to seafloor-hydrothermal sulfide deposits, central Arizona, USA. Earth Planet. Sci. Lett. 255, 243-256.

Sleep, N.H., Zahnle, K.J., Kasting, J.F., Morowitz, H.J., 1989. Annihilation of ecosystems by large asteroid impacts on the early Earth. Nature 342, 139-142.

Stanley, S.M., 1979. Macroevolution: process and product. Freeman, San Francisco.

Steed, P.R., Fillingame, R.H., 2008. Subunit *a* facilitates aqueous access to a membrane-embedded region of subunit *c* in *Escherichia coli* $F_1F_o$ ATP Synthase. J. Biol. Chem. 283, 12365-12372.

Steigmiller, S., Turina, P., Gräber, P., 2008. The thermodynamic $H^+$/ATP ratios of the $H^+$-ATP synthases from chloroplasts and *Escherichia coli*. Proc. Natl Acad. Sci. USA 105, 3745-3750.

Steinberg, I.Z., Oplatka, A., Katchalsky, A., 1966. Mechanochemical engines. Nature 210, 568-571.

Stepney, S., 2008. The neglected pillar of material computation. Physica D 237, 1157-1164.

Stöffler, D., Horneck, G., Ott, S., Hornemann, U., Cockell, C.S., Moeller, R., Meyer, C., De Vera, J.-P., Fritz, J., Artemieva, A., 2008. Experimental evidence for the potential impact ejection of viable microorganisms from Mars and Mars-like planets. Astrobiology 8, 585-588.

Sun, F.-J., Caetano-Anolles, G., 2008. The origin and evolution of tRNA inferred from phylogenetic analysis of structure. J. Mol. Evol. 66, 21-35.

Tajika, E., 2003. Faint young Sun and the carbon cycle: implications for the Proterozoic global glaciations, Earth Planet. Sci. Lett. 214, 443-453.

Tesla, N., 1889. Thermo magnetic motor. U.S. Patent 396121.

Teuscher, C., Nemenman, I., Alexander, F.J., 2008. Novel computing paradigms: Quo vadis? Physica D 237, v-viii,

Thauer, R.K., 1998. Biochemistry of methanogenesis: a tribute to Marjorie Stephenson. Microbiology 144, 2377-2406.

Todgham, A.E., Hoaglund, E.A., Hoffman, G.E., 2007. Is cold the new hot? Elevated ubiquitin-conjugated protein levels in tissues of Antarctic fish as evidence for cold-denaturation of proteins in vivo. J. Comp. Physiol. B 177, 857-866.

Toei, M., Gerle, C., Nakano, M., Tani, K., Gyobu, N., Tamakoshi, M., Sone, N., Yoshida, M., Fujiyoshi, Y., Mitsuoka, K., Yokoyama, K., 2007. Dodecamer rotor ring defines $H^+$/ATP ratio for ATP synthesis of prokaryotic V-ATPase from *Thermus thermophilus*. Proc. Natl Acad. Sci. USA 104, 20256-20261.

Tomasek, J.J., Brusilow, W.S.A., 2000. Stoichiometry of energy coupling by proton-translocating ATPases: a history of variability. J. Bioenerg. Biomemb. 32, 493-500.

Tomizaki, K.-Y., Mihara, H., 2007. Phosphate-mediated molecular memory driven by two different protein kinases as information input elements. J. Am. Chem. Soc. 129, 8345-8352.

Trinks, H., Schroeder, W., Biebricher, C.K., 2005. Ice and the origin of life. Origins Life Evol. Biosph. 35, 429-445.

Turing, A., 1937. On computable numbers, with an application to the Entscheidungsproblem. Proc. London Math. Soc. 2, 42, 230-265.

Turner, L., Caplan, S.R., Berg, H.C., 1996. Temperature-induced switching of the bacterial flagellar motor. Biophys. J. 71, 2227-2233.

Tuszynski, J.A., Hameroff, S., Sataric, M.V., Trpisova, B., Nip, M.L.A., 1995. Ferroelectric behaviour in microtubule dipole lattices: implications for information processing, signalling and assembly/disassembly. J. Theor. Biol. 174, 371-380.

Unger, R., Moult, J., 2006. Towards computing with proteins. Proteins 63, 53-64.

Valley, J.W., Peck, W.H., King, E.M., Wilde, S.A., 2002. A cool early Earth. Geology 30, 351-354.

Van der Pol, B., Van der Mark, J., 1928. The heartbeat considered as a relaxation oscillation, and an electrical model of the heart. Philos. Mag. Suppl. 6, 763-775.

Van Holde, K.E., 1980. The origins of life: a thermodynamic critique. In: Halverson, H.O., Van Holde K.E., (Eds), The Origins of Life and Evolution. Alan Liss, New York, pp. 31-46.

Vekshin, N.L., 1991. Thermal-model of coupling ATP synthesis with electron-transfer. Biofizika 36, 994-999.

Vellai, T., Vida, G., 1999. The origin of eukaryotes: the difference between prokaryotic and eukaryotic cells. Proc. R. Soc. Lond. B 266, 1571-1577.

Villaverde, J., Cladera, J., Hartog, A., Berden, J., Padros, E., Dunach, M., 1998. Nucleotide and $Mg^{2+}$ dependency of the thermal denaturation of mitochondrial $F_1$-ATPase. Biophys. J. 75, 1980-1988.

Vincent, O.D., Schwem, B.E., Steed, P.R., Fillingame, R.H., 2007. Fluidity of structure and swiveling of helices in the subunit $c$ ring of Escherichia coli ATP Synthase as revealed by cysteine-cysteine cross linking. J. Biol. Chem. 282, 33788-33794.

Vinogradov, A.D., 1999. Mitochondrial ATP Synthase: fifteen years later. Biochemistry (Moscow) 64, 1219-1229.

Vlassov, A.V., Kazakov, S.A., Johnston, B.H., Landweber, L.F., 2005. The RNA World on ice: a new scenario for the emergence of RNA information. J. Mol. Evol. 61, 264-273.

Von Neumann, J., 1956. Probabilistic logics and the synthesis of reliable organisms from unreliable computers. In: Shannon, C.E., McCarthy, J., (Eds.), Automata studies. Princeton University Press, pp. 43-98.

Von Smoluchowski, M., 1912. Experimentell nachweisbare, der üblichen Thermodynamik widersprechende Molekularphänome. Physik. Zeitschr. 13, 1069-1080.

Vorburger, T., Ebneter, J.Z., Wiedenmann, A., Morger, D., Weber, G., Diederichs, K., Dimroth, P., von Ballmoos, C., 2008. Arginine-induced conformational change in the $c$-ring/$a$-subunit interface of ATP Synthase. FEBS J. 275, 2137-2150.

Wang, G.M., Sevick, E.M., Mittag, E., Searles, D.J., Evans, D.J., 2002. Experimental demonstrations of violations of the second law of thermodynamics for small systems and short time scales. Phys. Rev. Letters 89, 0500601.

Wang, Z.Y., Freire, E., McCarthy, R.E., 1993. Influence of nucleotide binding site occupancy on the thermal stability of the $F_1$ portion of the chloroplast ATP Synthase. J. Biol. Chem. 268, 20785-20790.

Warren, B.A., 1981. Deep circulation of the world ocean. In: Warren B.A., (Ed.) Evolution of physical oceanography. Scientific surveys in honor of Henry Stommel. MIT Press, Cambridge, pp. 6-41.

Webre, D.J., Wolanin, P.M., Stock, J.B., 2003. Bacterial chemotaxis. Current Biology 13, R47-R49.

Wekjira, J.F., 2001. The VT1 shape memory alloy heat engine design. Thesis Virginia Polytechnic Institute, Blacksburg VA.

White, S.H., VonHeijne, G., 2008. How translocons select transmembrane helices. Annu. Rev. Biophys. 37, 23-42.

Wickner, W., Schekman, R., 2005. Protein translocation across biological membranes. Science 310, 1452-1456.

Wiegand, W.B., Snyder, J.W., 1934. The rubber pendulum, the Joule effect, and the dynamic stress strain curve. Inst. Rubber Ind. Trans. Proc. 10, 234-262.

Williams, R.J.P., 1961. Possible functions of chains of catalysts. J. Theor. Biol. 1, 1-17.

Woese, C., 1987. Bacterial evolution. Microbiol. Rev. 51, 221-271.

Wöhlisch, E., 1940. Muskelphysiologie vom Standpunkt der kinetische Theorie der Hochelastizität und der Entspannungshypothese des Kontraktionsmechanismus. Naturwissenschaften 28, 305-312.

Wolf, D.M., Fontaine-Bodin, L., Bischofs, I., Price, G., Keasling, J., Arkin, A.P., 2006. Memory in microbes: quantifying history-dependent behaviour in a bacterium. PLoS ONE 3, e1700 1-114.

Wunsch, C., Ferrari, R., 2004. Vertical mixing, energy, and the general circulation of the oceans. Annu. Rev. Fluid Mech. 36, 281-314.

Xiao, S., Shen, B., Zhou, C., Xie, G., Yuan, X., 2005. A uniquely preserved Ediacaran fossil with direct evidence for a quilted bodyplan. Proc. Natl Acad. Sci. USA 102, 10227-10232.

Young, G.M., von Brunn, V., Gold, D.J.C., Minter, W.E.L., 1998. Earth's oldest reported glaciation; physical and chemical evidence from the Archean Mozaan Group (~2.9 Ga) of South Africa. J. Geol. 106, 523-538.

Zhang, K., Schubert, G., 2000. Magnetohydrodynamics in rapidly rotating spherical shells. Annu. Rev. Fluid Mech. 32, 411-445.

Zhang, Y., Chen, J., 2008. Investigation on a temporal asymmetric oscillation temperature ratchet. Physica A 387, 3443-3448.

## Appendix

In March of this year (2009) I received a message of rejection from the journal to which I had submitted the paper. I first give the comments of the reviewer as whole. Thereafter I react to the specific comments in detail: this reaction has not been sent to the journal concerned.

```
In the paper, the author presents scenarios of how heat engines
might have contributed to the evolution of animals. While the
topic of the paper, biological heat engines, is fascinating and
deserves attention, the paper is in my opinion by form and
content not suitable for publication in [name of the journal]. The
manuscript looks to me like a collection of ideas centered on
the topic of heat engine. Although some of these ideas might be
plausible, it is far from presenting a coherent and plausible
scenario for the emergence of animals from heat engines.

Specific comments:

The paper is about 70 pages long and contains 21 figures and is
just the first part of a series of papers. It gives an unusually
subjective and speculative account on the origin of life, and
excessively discusses side-points that do not contribute to the
main argument of the author. This clearly goes much beyond what
is normally published in [name of the journal]. In my opinion, a book
might be a more suitable format for this work.

Although research on the origin of life necessarily contains
speculative elements, there a numerous findings that are more or
less accepted. A recent overview is, for example, given in
"Introduction: how and when did microbes change the world?"
(Thomas Cavalier-Smith, Martin Brasier & T. Martin Embley,
Philosophical Transactions of The Royal Society B 361 (2006)
pages: 845-850.) The paper fails to discuss how the scenarios
fit together with established findings from origin of live
research. It fails not give reference to many important papers
in this field.

The main argument for heat engines put forward by the author is
that thermal gradients provided by hydrothermal vents were the
only reliable energy source during global glaciation events. I
am not convinced by this point. Hydrothermal vents are not only
a source of thermal gradients but also of diverse chemical
substrates for energy-production.

I cannot follow the discussion of the "flagellar computer". It
is questionable highly whether the flagellum can do calculations
comparable to a Turing machine. Moreover, I cannot see that
flagellums actually do such calculations in reality. Biological
computation is mainly performed by biochemical signal
transduction. Chemotaxis, for example, is a well-understood
process that does not rely on the notion of the flagellum as a
Turing machine.

The title of the manuscript seems misleading for the scenarios
presented so far. What specific events in the evolution of
animals are exactly referred to? So far, the author discusses
the emergence of flagellated organisms. Thus the scenario so far
applies to eukaryotes and most prokaryotes. It is unclear why
heat engines using convection are associated with photosynthesis
```

```
while sessile organisms using the mechanisms described in the
paper are associated with the emergence of animals.

In the current version it is difficult to reconstruct the order
and timing of the events described in the manuscript. Here, a
figure might be helpful. Such a figure could also link to
established events in the origin of life and would be helpful to
compare the scenario with other scenarios put forward in the
field.
```

My reaction:

```
The paper is about 70 pages long and contains 21 figures and is
just the first part of a series of papers. It gives an unusually
subjective and speculative account on the origin of life, and
excessively discusses side-points that do not contribute to the
main argument of the author. This clearly goes much beyond what
is normally published in [name of the journal]. In my opinion, a book
might be a more suitable format for this work.
```

The reviewer calls the paper long, unusually subjective and speculative. On its length I agree, although in the journal far less pages than 70 would be filled. One of the reasons for selecting the journal was that it has no prescribed paper length. On the unusual called subjectivity and speculation I would answer that the paper is based on the application of basic, well verified processes known to physics, chemistry and engineering. The application to biology is new. The criterion for calling a paper subjective and speculative differs strongly with the discipline. In physics and engineering proposing a new type of engine would not as quickly be called subjective as in biology. All proposed mechanisms are well supported by arguments. A book may indeed be a more suitable format.

```
Although research on the origin of life necessarily contains
speculative elements, there a numerous findings that are more or
less accepted. A recent overview is, for example, given in
"Introduction: how and when did microbes change the world?"
(Thomas Cavalier-Smith, Martin Brasier & T. Martin Embley,
Philosophical Transactions of The Royal Society B 361 (2006)
pages: 845-850.) The paper fails to discuss how the scenarios
fit together with established findings from origin of live
research. It fails not give reference to many important papers
in this field.
```

I am familiar with Cavaliers-Smith's papers, and many papers of other authors. I must restrict myself, however, because this paper is, admittedly, already quite long.

```
The main argument for heat engines put forward by the author is
that thermal gradients provided by hydrothermal vents were the
only reliable energy source during global glaciation events. I
am not convinced by this point. Hydrothermal vents are not only
a source of thermal gradients but also of diverse chemical
substrates for energy-production.
```

The paper explains clearly that chemical substrates can be energy source only when a chemical disequilibrium is present. But a redox disequilibrium will quickly vanish in the absence of photosynthesis, and fermentable substrates, if present at all, would require a large set of enzymes.

```
I cannot follow the discussion of the "flagellar computer". It
is questionable highly whether the flagellum can do calculations
comparable to a Turing machine. Moreover, I cannot see that
flagellums actually do such calculations in reality. Biological
computation is mainly performed by biochemical signal
transduction. Chemotaxis, for example, is a well-understood
process that does not rely on the notion of the flagellum as a
Turing machine.
```

Chemotaxis is not well-understood, and has not been completely clarified. Any information processing can be called a computer. An error in the text was present though: no distinction was made between a 'Turing machine' and a 'Universal Turing Machine'. This has been corrected.

```
The title of the manuscript seems misleading for the scenarios
presented so far. What specific events in the evolution of
animals are exactly referred to? So far, the author discusses
the emergence of flagellated organisms. Thus the scenario so far
applies to eukaryotes and most prokaryotes. It is unclear why
heat engines using convection are associated with photosynthesis
while sessile organisms using the mechanisms described in the
paper are associated with the emergence of animals.
```

The paper is indeed meant as a first paper out of a series of papers on the origin of animals. Previous studies of mine explain the association between convection, with its associated light-dark cycling and photosynthesis.

```
In the current version it is difficult to reconstruct the order
and timing of the events described in the manuscript. Here, a
figure might be helpful. Such a figure could also link to
established events in the origin of life and would be helpful to
compare the scenario with other scenarios put forward in the
field.
```

An additional figure would indeed help, but would make the paper even longer. In hindsight I agree that a first figure with a chronology would have been an improvement.

The paper is indeed long. The arguments given could sometimes indeed better be formulated. Nevertheless, I consider it a pity that the reviewer could not recognize the many new explanations and ideas presented and did not balance them with its blemishes, but instead focused on the latter, and decided for rejection.